%% file: main.tex
\documentclass[sigconf]{acmart}

\input{package}
\input{commands}
\AtBeginDocument{%
  \providecommand\BibTeX{{%
    \normalfont B\kern-0.5em{\scshape i\kern-0.25em b}\kern-0.8em\TeX}}}
    
 \newif\ifcameraready
\camerareadyfalse

\copyrightyear{2023}
\acmYear{2023}
\setcopyright{acmlicensed}\acmConference[ICS '23]{2023 International Conference on Supercomputing}{June 21--23, 2023}{Orlando, FL, USA}
\acmBooktitle{2023 International Conference on Supercomputing (ICS '23), June 21--23, 2023, Orlando, FL, USA}
\acmPrice{15.00}
\acmDOI{10.1145/3577193.3593719}
\acmISBN{979-8-4007-0056-9/23/06}

\begin{document}

\rtitle{\mech: Spatial Acceleration for Efficient and Scalable \\Horizontal Diffusion Weather Stencil Computation}

\title[\mech]{\mech: Spatial Acceleration for Efficient and Scalable \\Horizontal Diffusion Weather Stencil Computation}




\author{\vspace{-0.3cm}Gagandeep Singh$^{a,b}$ \hspace{0.5cm}   Alireza Khodamoradi$^{a}$ \hspace{0.5cm}  Kristof Denolf$^a$  \hspace{0.5cm}  Jack Lo$^a$  \\Juan G{\'o}mez-Luna$^b$ \hspace{0.5cm}  Joseph Melber$^a$ \hspace{0.5cm}  Andra Bisca$^a$\hspace{0.5cm}  \\ Henk Corporaal$^c$ \hspace{0.5cm} Onur Mutlu$^b$ 
\\ \normalsize $^a$AMD Research  \hspace{1cm}  $^b$ETH Z{\"u}rich \hspace{1cm} $^c$Eindhoven University of Technology\vspace{0.4cm}
}

\renewcommand{\shortauthors}{G. Singh, et al.}

\renewcommand{\authors}{Gagandeep Singh, Alireza Khodamoradi, Kristof Denolf, Jack Lo, Juan G{\'o}mez-Luna, Joseph Melber, Andra Bisca, Henk Corporaal, Onur Mutlu}

\input{sections/00-abstract}
\vspace{10pt}

\ccsdesc[500]{Hardware~Hardware-software codesign}
\ccsdesc[500]{Hardware~Emerging architectures}
\ccsdesc[500]{Computer systems organization~Parallel architectures}
\ccsdesc[500]{Computer systems organization~Data flow architectures}
\keywords{spatial computing systems, dataflow architectures, high-performance computing,  hybrid systems, weather prediction, stencil computation\gom{, memory access patterns, climate modeling}}

\settopmatter{printfolios=true} 
\maketitle
\pagenumbering{arabic} 


\begin{sloppypar}
\input{sections/01-introduction}

\input{sections/02-background}

\input{sections/03-implementation}

\input{sections/04-evaluation}

\input{sections/05-discussion}

\input{sections/06-related}

\input{sections/07-conclusion}
\end{sloppypar}

\begin{acks}
\gom{We thank \gomm{the} anonymous reviewers of ICS 2023 for their feedback. We thank the SAFARI
Research Group members for valuable feedback \gomm{and the stimulating
intellectual \gomm{and scientific} environment they provide}.} SAFARI Research Group acknowledges the generous gifts of their industrial partners. This research was partially supported by the Semiconductor Research Corporation.
\end{acks}
{
\balance
\bibliographystyle{IEEEtran}
\bibliography{references}
}


\end{document}

%% file: package.tex
\usepackage{soul} 
\usepackage{geometry}
\usepackage{array}
\usepackage{fancyhdr}
\usepackage[linesnumbered,ruled]{algorithm2e}
\usepackage{algpseudocode}

\usepackage{lscape}
\usepackage{xcolor}
\usepackage{textcomp}
\usepackage{tikz}

\usepackage{enumitem}
\usepackage{pythonhighlight}

\usepackage{balance}
\usepackage{tikz}
\usepackage[rightcaption]{sidecap} 

\soulregister\cite7
\soulregister\ref7
\soulregister\pageref7
\usepackage[acronym]{glossaries}



%% file: commands.tex
\input{tikz_hdiff}

\input{tikz_layout}
\newcommand{\head}[1]{{\noindent\textbf{#1.}\xspace}} 
\newcommand{\aie}{AIE\xspace}
\newcommand{\aies}{AI Engines\xspace}
\newcommand{\hdiff}{\texttt{hdiff}\xspace}

\definecolor{NavyBlue}{rgb}{0.19, 0.55, 0.91}
\newcolumntype{H}{>{\setbox0=\hbox\bgroup}c<{\egroup}@{}}

\newcommand{\jacOneD}{\texttt{{j}acobi-1d}\xspace} 
\newcommand{\jacTwoDT}{\texttt{{j}acobi-2d-3pt}\xspace} 
\newcommand{\jacTwoDN}{\texttt{{j}acobi-2d-9pt}\xspace} 
\newcommand{\sei}{\texttt{{s}eidel-2d}\xspace} 

\newcommand{\saie}{single-AIE\xspace} 
\newcommand{\maie}{multi-AIE\xspace} 
\newcommand{\daie}{dual-AIE\xspace} 
\newcommand{\taie}{tri-AIE\xspace} 
 
\newcommand{\srs}{\texttt{srs()}\xspace} 
\newcommand{\lap}{\texttt{Laplacian}\xspace} 
\newcommand{\flx}{\texttt{flux}\xspace} 

\newcommand{\MYhref}[3][blue]{\href{#2}{\color{#1}{#3}}}%

    \makeatletter
    \g@addto@macro{\normalsize}{%
      \setlength{\abovedisplayskip}{2pt plus 1pt minus 1pt}
      \setlength{\belowdisplayskip}{2pt plus 1pt minus 1pt}
      \setlength{\abovedisplayshortskip}{0pt}
      \setlength{\belowdisplayshortskip}{0pt}
      \setlength{\intextsep}{2pt plus 1pt minus 1pt}
      \setlength{\textfloatsep}{4pt plus 1pt minus 1pt}
      \setlength{\skip\footins}{5pt plus 1pt minus 1pt}}
    \makeatother

\newcommand{\fig}[1]{{Figure~#1}\xspace} 
\newcommand{\taway}[2]{{\noindent\textbf{\gom{Key Takeaway} \##1. #2.}\xspace}} 
\definecolor{dred}{rgb}{0.75,0.00, 0.00}
\definecolor{dpink}{rgb}{0.75,0.00, 0.75}
\definecolor{ddpink}{rgb}{1.0, 0.20, 1.0}
\definecolor{dgreen}{rgb}{0.0, 0.60, 0.30}
\definecolor{dblack}{rgb}{0.00, 0.0, 0.00}
\definecolor{dblue}{rgb}{0.00, 0.00, 0.75}
\definecolor{feedb}{rgb}{0.75, 0.00, 0.75}

\definecolor{aqua}{rgb}{0.0, 0.90, 1.0}

\newcommand{\gs}[1]{{\color{black}#1}}

\newcommand{\ics}[1]{{\color{black}#1}}
\newcommand{\gcam}[1]{{\color{black}#1}}

\newcommand{\gom}[1]{{\color{black}#1}} 
\newcommand{\gomm}[1]{{\color{black}#1}} 
\newcommand{\gommm}[1]{{\color{black}#1}}

\balance
  
\newcommand{\mech}{SPARTA\xspace}
\newcommand{\hdiffcpu}{17.1$\times$\xspace}
\newcommand{\hdifffpga}{2.1$\times$\xspace}
\newcommand{\hdiffgpu}{1.2$\times$\xspace}
\definecolor{dodgerblue}{rgb}{0.12, 0.56, 1.0}

\newcommand*\circled[1]{\tikz[baseline=(char.base)]{
            \node[shape=circle,draw,inner sep=0pt,fill=black, text=white] (char) {#1};}}
\newcommand*\circledWhite[1]{\tikz[baseline=(char.base)]{
            \node[shape=circle,draw,inner sep=0pt,fill=white, text=black] (char) {#1};}}

\makeatletter

\newcommand{\disableAcronymHyperlink}{%
  \def\AC@hyperlink##1##2{##2}%
  \def\AC@hyperref[##1]##2{##2}%
  \def\AC@hypertarget##1##2{##2}%
  \def\AC@phantomsection{}%
}
\makeatother

\SetCommentSty{mycommfont}

%% file: tikz_hdiff.tex
\newcommand{\myGlobalTransformation}[2]
{
    \pgftransformcm{1}{0}{0.4}{0.5}{\pgfpoint{#1cm}{#2cm}}
}

\newcommand{\gridThreeD}[3]
{
    \begin{scope}
         \myGlobalTransformation{#1}{#2};
        \draw [#3] grid (5,5);
        \def\opa{40}
        \fill[ gray!\opa] (2,2) -- +(0, 1) -- +(1, 1) -- +(1,0) -- cycle;
        \fill[ gray!\opa] (3,3) -- +(0, 1) -- +(1, 1) -- +(1,0) -- cycle;
        \fill[ gray!\opa] (2,4) -- +(0, 1) -- +(1, 1) -- +(1,0) -- cycle;
        \fill[ gray!\opa] (1,1) -- +(0, 1) -- +(1, 1) -- +(1,0) -- cycle;
        \fill[ gray!\opa] (1,3) -- +(0, 1) -- +(1, 1) -- +(1,0) -- cycle;
        \fill[ gray!\opa] (0,2) -- +(0, 1) -- +(1, 1) -- +(1,0) -- cycle;
        \fill[ gray!\opa] (4,2) -- +(0, 1) -- +(1, 1) -- +(1,0) -- cycle;
        \fill[ gray!\opa] (3,1) -- +(0, 1) -- +(1, 1) -- +(1,0) -- cycle;
        \fill[ gray!\opa] (2,0) -- +(0, 1) -- +(1, 1) -- +(1,0) -- cycle;
    \end{scope}
}

\newcommand{\gridThreeDSecond}[3]
{
    \begin{scope}
        \myGlobalTransformation{#1}{#2};
        \draw [#3] grid (3,3);
         \def\opa{40}
        \fill[ gray!\opa] (0,1) -- +(0, 1) -- +(1, 1) -- +(1,0) -- cycle;
         \def\opa{40}
        \fill[ gray!\opa] (1,0) -- +(0, 1) -- +(1, 1) -- +(1,0) -- cycle;
        \fill[ gray!\opa] (1,2) -- +(0, 1) -- +(1, 1) -- +(1,0) -- cycle;
        \fill[ gray!\opa] (2,1) -- +(0, 1) -- +(1, 1) -- +(1,0) -- cycle;
    \end{scope}
}

\newcommand{\gridThreeDThird}[3]
{
    \begin{scope}
        \myGlobalTransformation{#1}{#2};
        \def\opa{40}
        \draw [#3] grid (1,1);
        \fill[ gray!\opa] (0,0) -- +(0, 1) -- +(1, 1) -- +(1,0) -- cycle;
    \end{scope}
}
\tikzstyle myBG=[line width=3pt,opacity=1.0]


%% file: tikz_layout.tex
\newcommand{\laplines}{
    \def\len{32}
    \def\lenright{24}
     \def\lenleft{8}
      \def\lenleftf{2}
       \def\w{0.02cm}
       \def\lapht{-1.9}
    { 
     \def\col{violet}
     \def\laploc{15.5}
      \draw[\col,line width=\w] (\len/2+0.5,0)--(\laploc,\lapht);
      \draw[\col,line width=\w] (\len/2+1+0.5,0)--(\laploc,\lapht);
      \draw[\col,line width=\w] (\len/2-1+0.5,0)--(\laploc,\lapht);
      \draw[\col,line width=\w] (\lenright+0.5,0)--(\laploc,\lapht);
      \draw[\col,line width=\w] (\lenleft+0.5,0)--(\laploc,\lapht);
    }
    { 
     \def\col{blue}
     \def\laploc{13.5}
    \draw[\col,line width=\w] (\len/2+0.5,0)--(\laploc,\lapht);
    \draw[\col,line width=\w] (\len/2-1+0.5,0)--(\laploc,\lapht);
    \draw[\col,line width=\w] (\len/2-2+0.5,0)--(\laploc,\lapht);
    \draw[\col,line width=\w] (\lenright-1+0.5,0)--(\laploc,\lapht);
    \draw[\col,line width=\w] (\lenleft-1+0.5,0)--(\laploc,\lapht);
    }
    
      { 
     \def\col{red}
     \def\laploc{17.5}
    \draw[\col,line width=\w] (\len/2+0.5,0)--(\laploc,\lapht);
    \draw[\col,line width=\w] (\len/2+1+0.5,0)--(\laploc,\lapht);
    \draw[\col,line width=\w] (\len/2+2+0.5,0)--(\laploc,\lapht);
    \draw[\col,line width=\w] (\lenright+1+0.5,0)--(\laploc,\lapht);
    \draw[\col,line width=\w] (\lenleft+1+0.5,0)--(\laploc,\lapht);
    }
    
     { 
     \def\col{brown}
     \def\laploc{9.5}
    \draw[\col,line width=\w] (\lenleft-1+0.5,0)--(\laploc,\lapht);
    \draw[\col,line width=\w] (\lenleft+0.5,0)--(\laploc,\lapht);
    \draw[\col,line width=\w] (\lenleft+1+0.5,0)--(\laploc,\lapht);
    \draw[\col,line width=\w] (\lenleftf+0.5,0)--(\laploc,\lapht);
    \draw[\col,line width=\w] (\len/2+0.5,0)--(\laploc,\lapht);
    }
    { 
     \def\col{orange}
     \def\laploc{21.5}
    \draw[\col,line width=\w] (\lenright-1+0.5,0)--(\laploc,\lapht);
    \draw[\col,line width=\w] (\lenright+0.5,0)--(\laploc,\lapht);
    \draw[\col,line width=\w] (\lenright+1+0.5,0)--(\laploc,\lapht);
    \draw[\col,line width=\w] (\lenrightf+0.5,0)--(\laploc,\lapht);
    \draw[\col,line width=\w] (\len/2+0.5,0)--(\laploc,\lapht);
    }
}

\newcommand{\fluxlines}{
    \def\len{32}
    \def\lenright{24}
     \def\lenleft{8}
      \def\lenleftf{2}
      \def\w{0.02cm}
       \def\fluxht{-5.9}
     { 
     \def\col{pink}
     \def\lapij{15.5}
     \def\fluxloc{13.5}
    
      \draw[\col,line width=\w] (\lapij,-4)--(\fluxloc,\fluxht);
      \draw[\col,line width=\w] (\lapij-2,-4)--(\fluxloc,\fluxht);
      \draw[dashed,\col,line width=\w] (\len/2+0.5,0).. controls +(down:7mm) and +(right:25mm) .. (\fluxloc,\fluxht);
      \draw[dashed,\col,line width=\w] (\len/2-1+0.5,0).. controls +(down:7mm) and +(left:25mm) .. (\fluxloc,\fluxht);
      }
      
      { 
     \def\col{olive}
     \def\lapij{15.5}
     \def\fluxloc{17.5}
      
      \draw[\col,line width=\w] (\lapij,-4)--(\fluxloc,\fluxht);
      \draw[\col,line width=\w] (\lapij+2,-4)--(\fluxloc,\fluxht);
      \draw[dashed,\col,line width=\w] (\len/2+0.5,0).. controls +(down:7mm) and +(right:15mm) .. (\fluxloc,\fluxht);
      \draw[dashed,\col,line width=\w] (\len/2+1+0.5,0).. controls +(down:7mm) and +(left:15mm) .. (\fluxloc,\fluxht);
      }
     
        { 
     \def\col{teal}
     \def\lapij{15.5}
     \def\fluxloc{11.5}
      \draw[\col,line width=\w] (\lapij,-4)--(\fluxloc,\fluxht);
      \draw[\col,line width=\w] (\lapij-6,-4)--(\fluxloc,\fluxht);
      \draw[dashed,\col,line width=\w] (\len/2+0.5,0).. controls +(down:7mm) and +(right:15mm) .. (\fluxloc,\fluxht);
      \draw[dashed,\col,line width=\w] (\lenleft+0.5,0).. controls +(down:7mm) and +(left:15mm) .. (\fluxloc,\fluxht);
      }
      
         { 
     \def\col{cyan}
     \def\lapij{15.5}
     \def\fluxloc{19.5}
    
      \draw[\col,line width=\w] (\lapij,-4)--(\fluxloc,\fluxht);
      \draw[\col,line width=\w] (\lapij+6,-4)--(\fluxloc,\fluxht);
      \draw[dashed,\col,line width=\w] (\len/2+0.5,0).. controls +(down:7mm) and +(right:15mm) .. (\fluxloc,\fluxht);
      \draw[dashed,\col,line width=\w] (\lenright+0.5,0).. controls +(down:7mm) and +(left:15mm) .. (\fluxloc,\fluxht);
      }
}

\newcommand{\outlines}{
    \def\len{32}
    \def\lenright{24}
     \def\lenleft{8}
      \def\lenleftf{2}
       \def\outht{-9.9}
      
     \def\col{violet}
     \def\w{0.03cm}
     \draw[\col,line width=\w] (11.5,-8)--(\len/2+0.5,\outht);
     \draw[\col,line width=\w] (13.5,-8)--(\len/2+0.5,\outht);
     \draw[\col,line width=\w] (17.5,-8)--(\len/2+0.5,\outht);
     \draw[\col,line width=\w] (19.5,-8)--(\len/2+0.5,\outht);
     \draw[dashed,\col,line width=\w] (\len/2+0.5,0).. controls +(down:7mm) and +(left:15mm) .. (\len/2+0.5,\outht+0.2);
     
     }

\newcommand\inputLayout[1]{

     \def\len{32}
     \def\lenright{24}
     \def\lenleft{8}
     \def\lenrightf{30}
     
     \def\lenleftf{2}
      
     \foreach \a in {0,...,\len} 
     {\draw[fill=white](\a,0) rectangle ++(1,2);}
      \node at (3,2.5)[rotate=0] {{Input Array}};
      \node at (4,-3)[rotate=0] {{Laplace Results}};
      \node at (28,-1)[rotate=0] {\textit{Laplace}};
      \node at (28,-1.8)[rotate=0] {\textit{Stencil}};
      \node at (26,-5)[rotate=0] {\textit{Flux}};
      \node at (26,-5.85)[rotate=0] {\textit{Stencil}};
      \node at (3,-7)[rotate=0] {{Flux Results}};
       \node at (3,-9.5)[rotate=0] {{Write Array}};
      
     \draw[fill=white](8,-4) rectangle ++(16,2);
     \draw[fill=white](10,-8) rectangle ++(12,2);
    \foreach \a in {0,...,\len} 
    {\draw[fill=white](\a,-12) rectangle ++(1,2);}
       
          \foreach \ij in {\len/2,\len/2-1,\len/2-2,\len/2+1,\len/2+2} {
           \draw[fill=gray!50](\ij,0) rectangle ++(1,2);
        }
       
         \foreach \ipj in {\lenright-1,\lenright,\lenright+1} {
           \draw[fill=gray!50](\ipj,0) rectangle ++(1,2);
            }
          \foreach \imj in {\lenleft-1,\lenleft,\lenleft+1} {
           \draw[fill=gray!50](\imj,0) rectangle ++(1,2);
            }
            \draw[fill=gray!50](\lenrightf,0) rectangle ++(1,2);
            \draw[fill=gray!50](\lenleftf,0) rectangle ++(1,2);

       \foreach \a in {9,13,15,17,21} {
          \draw[fill=gray!20](\a,-4) rectangle ++(1,2);
      }

       \foreach \a in {11,13,17,19} {
          \draw[fill=gray!20](\a,-8) rectangle ++(1,2);
      }
      
       \draw[fill=gray!50](\len/2,-12) rectangle ++(1,2);

    }

%% file: sections/00-abstract.tex
\begin{abstract}
Fast and accurate climate simulations and weather predictions are critical for understanding and preparing for the impact of climate change. 
Real-world climate and weather simulations involve the \gom{use} of complex compound stencil kernels, which are composed of a combination of different stencils. 
Horizontal diffusion is one such important compound stencil found in many 
climate and weather prediction models. 
Its computation involves a large amount of data \gom{access and} manipulation 
that leads to two main issues on current computing systems. First, \gom{such compound stencils} 
have high \gom{memory} bandwidth demands as \gom{they} require large amounts of data \gom{access}. Second,  compound stencils have complex
data access patterns and poor data locality, as  the memory access pattern is typically \gom{irregular with low arithmetic intensity.} 
As a result, state-of-the-art CPU and GPU implementations suffer from limited performance and high energy consumption. 
Recent works propose using FPGAs as an alternative to traditional CPU and GPU-based systems to accelerate weather stencil kernels. However, we observe that stencil computation cannot leverage the bit-level flexibility available on an FPGA because of its complex memory access patterns, 
 leading to high \gom{hardware} resource utilization and low peak performance. 

We introduce \mech,  a novel spatial accelerator for horizontal diffusion weather stencil
computation.  
We exploit the two-dimensional spatial architecture  to efficiently accelerate \gcam{the} horizontal diffusion stencil  
by designing the first scaled-out spatial accelerator using \gom{the} MLIR (Multi-Level Intermediate Representation) compiler framework. We evaluate \gom{\mech} on a real cutting-edge AMD-Xilinx Versal AI Engine (AIE) spatial architecture. Our real-system evaluation results demonstrate that \mech outperforms state-of-the-art CPU, GPU,  and FPGA implementations by \hdiffcpu, \hdiffgpu, and \hdifffpga 
, respectively. Compared to the  most  energy-efficient design on an HBM-based FPGA, \mech provides 2.43$\times$ higher energy efficiency.  Our results reveal that balancing workload across the available processing resources is crucial in achieving high performance on spatial architectures.   We also implement and evaluate five elementary stencils that are commonly used as benchmarks for stencil computation research. 
We freely open-source all our implementations to aid future research in stencil computation and spatial computing systems at \MYhref[cyan]{https://github.com/CMU-SAFARI/SPARTA}{https://github.com/CMU-SAFARI/SPARTA}. 





\end{abstract}

%% file: sections/01-introduction.tex
\section{Introduction}
Climate simulations and weather predictions are essential for understanding and preparing for the impacts of climate change.  The ability to make fast and accurate predictions is one of the biggest challenges meteorologists face to ensure effective and timely responses~\cite{bauer2021digital,hausfather2020evaluating,slingo2022ambitious,sillmann2017understanding,npg-30-13-2023,balaji2022general,hu2021efficient,dance2019improvements,hu2023progress,dueben2018challenges,pyle2021domain}. Researchers use mathematical models 
to simulate and analyze the behavior of the Earth's ecosystem. These models are based on fundamental physical and chemical principles and are continuously refined to improve their accuracy~\cite{bonaventura2000semi,cosmo_knl}.
The Consortium for Small-Scale
Modeling (COSMO)~\cite{doms1999nonhydrostatic} is one such model that is extensively used in weather forecasting, numerical weather prediction, and climate studies.

The main computational pipeline of COSMO\gom{,} called the \emph{dycore} or the dynamical core\gom{,} consists of compound stencil kernels that operate on a three-dimensional grid~\cite{gysi2015modesto,cosmo_knl}. 
Horizontal
diffusion (\hdiff) is a fundamental kernel found in the dynamical core  of  COSMO and other regional and global climate and weather prediction models~\cite{gysi2015modesto,cosmo_knl,de2021stencilflow}, including the European Centre for Medium-Range Weather Forecasts (ECMWF) model~\cite{palmer1990european} and the Global Forecast System (GFS) model~\cite{mcclung2016global}. It is also used in climate models, such as the Community Earth System Model (CESM)~\cite{hurrell2013community} and the Model for Interdisciplinary Research on Climate (MIROC)~\cite{watanabe2011miroc}.  The main use of horizontal diffusion is to help reduce the impact of small-scale errors in the model, such as those caused by the limited resolution of the model or by the effects of small-scale atmospheric processes that are not well represented by the model~\cite{daley2005horizontal}. By applying horizontal diffusion, the model can better represent the large-scale flow of the atmosphere and produce more accurate forecasts~\cite{skamarock2008time}.  

Recent works  propose the use of FPGAs~\cite{narmada,singh2020nero, singh2022designing,singh2022accelerating,singh2019low,de2021stencilflow,singh2021fpga,gagan_phd_thesis_2021} as an alternative to
traditional CPU and GPU-based systems in accelerating such weather stencil-based workloads due to FPGA's \gom{customizability to computation patterns}. However, we observe three main issues. First, compound
stencil computation cannot leverage the bit-level flexibility available on an FPGA because of its complex memory
access patterns, leading to high \gom{hardware} resource utilization. 
Second, the difficulty of designing and implementing efficient algorithms on FPGAs 
 has been a\gom{n impediment against} broad, high-volume adoption. Third, stencils in weather and climate simulations, such as horizontal diffusion, can be parameterized using different coefficients, which would entail designing a different hardware accelerator for each parameterization scheme.  

\begin{figure}[h]
  \centering
  \includegraphics[width=1\linewidth,trim={0cm 0cm 0cm 0cm},clip]{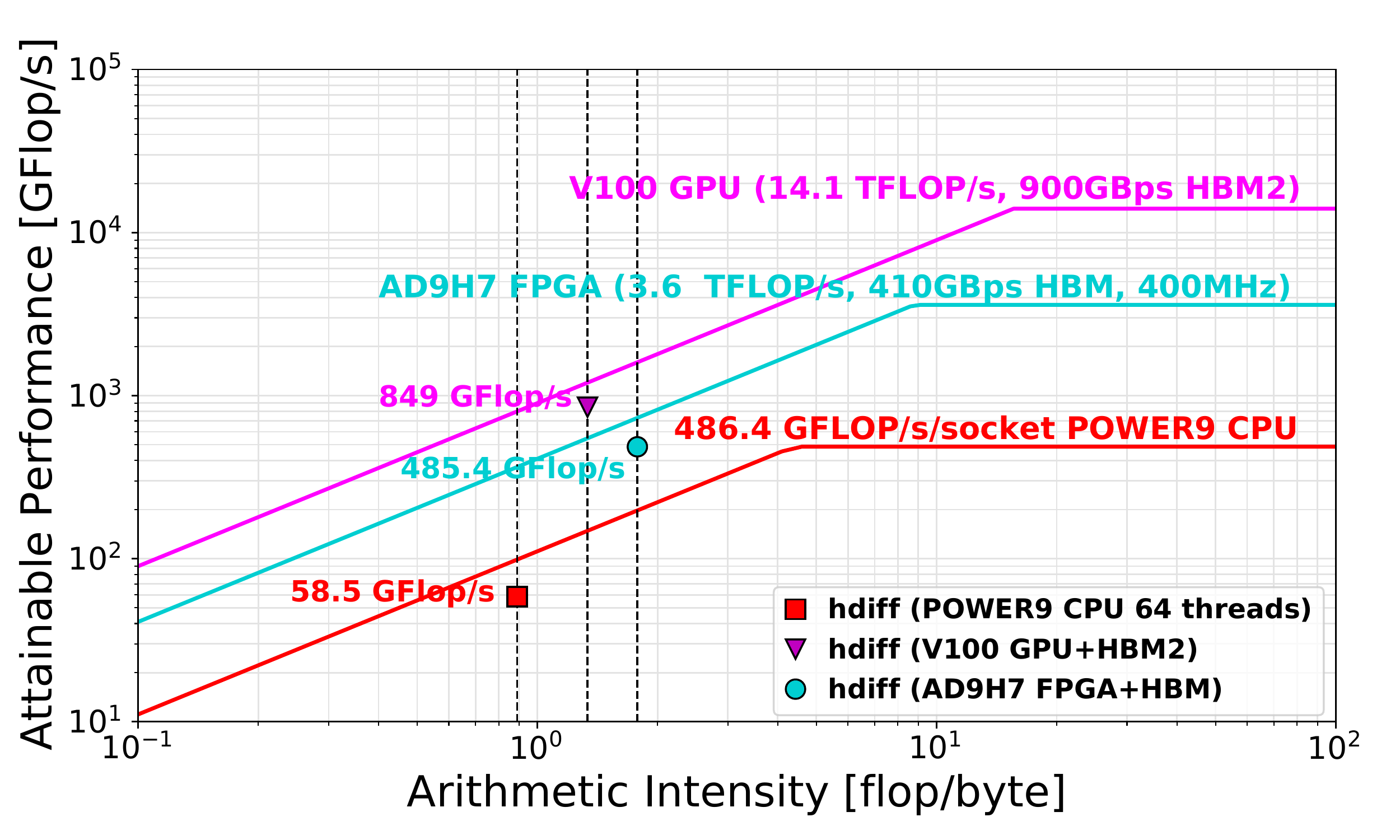}
     \caption{Roofline~\cite{williams2009roofline} for POWER9 (1-socket) showing horizontal diffusion {(\texttt{hdiff})} kernel for 64-thread 
    implementation. 
    The plot also shows the rooflines 
    of HBM-based NVIDIA V100 GPU~\cite{v100} and AD9H7 FPGA~\cite{ad9h7} with peak DRAM bandwidth. 
\label{fig:roof_hdiff} 
}
 \end{figure}

Figure~\ref{fig:roof_hdiff} shows roofline plots~\cite{williams2009roofline} for an IBM server-grade POWER9 CPU-\gomm{based}~\cite{POWER9} and NVIDIA V100 GPU~\cite{v100} \gom{from the state-of-the-art \hdiff CPU-~\cite{singh2020nero} and GPU-based~\cite{de2021stencilflow} implementations, respectively}. It also shows the roofline for an HBM-based AD9H7 FPGA~\cite{ad9h7} used in the state-of-the-art \hdiff implementation work~\cite{singh2020nero}. We make the following two major observations. First, \hdiff cannot achieve peak performance on a given architecture and often results in only 6.1\%-13.5\% \gom{of the peak performance} \gom{(i.e.,  the percentage of achieved peak roofline performance)} on current multi-core, GPU, and FPGA-based architectures. \gcam{Its} performance is dominated by memory-bound operations with
unique irregular memory access patterns and low arithmetic intensity that often results in poor performance on current systems.  Second, \hdiff is not able to leverage the bit-level programmability of an FPGA because of complex memory access patterns leading to high resource usage as
\gom{it} spends a large fraction of time on irregular memory operations.  

Recently, several spatial architectures, such as AMD-Xilinx Versal AI Engine~\cite{vissers2019versal}, Groq Tensor Streaming Processor~\cite{abts2020think}, Intel’s Configurable Spatial Accelerator ~\cite{morgan2018intel}, and Cerebras Deep Neural Network Accelerator~\cite{la2020cerebras}, have been proposed to accelerate machine learning-based workloads~\cite{zhuang2023charm,singh2022framework}. These architectures
consist of an array of small processing cores that are interconnected through a highly configurable network. The primary focus of these spatial architectures has been to provide a more efficient and effective way to accelerate machine learning workloads. However, we observe that spatial architectures can overcome the limitations of accelerating stencil computations on traditional architectures \gom{due to} three reasons. 
First, unlike FPGAs, which offer bit-level flexibility, spatial architecture\gom{s} provide coarse-grained flexibility that allows for tailoring of the dataflow to optimize data movement.  Second, the two-dimensional layout of \gom{spatial} architectures maps well to processing multi-dimensional grids, making them well-suited for stencil computation.  Third, the dataflow design of \gom{spatial architectures} provides an intuitive way to take advantage of both spatial and temporal locality  in iterative stencil processing by pipelining different timesteps.  Therefore, \gom{spatial} architectures have the potential to enable \gom{highly} efficient processing of weather stencil computation, leading to improved performance and scalability.

\textbf{Our goal} is to mitigate the
performance bottleneck of memory-bound weather stencil computation\gom{s} \gom{by taking advantage of the characteristics of}       spatial computing systems.

To this end, we introduce  \mech, a novel spatial accelerator for horizontal diffusion stencil computation. We evaluate \mech on \gom{the} real cutting-edge AMD-Xilinx Versal~\cite{vissers2019versal} spatial architecture. This architecture offers coarse-grained acceleration through the implementation of Vector VLIW (Very Long Instruction Word)~\cite{fisher1983very} processing cores, known as AI Engine\gom{s} (AIE\gom{s}).  We tailor the horizontal diffusion algorithm to \gom{fit well into the} spatial architecture datapath that optimizes data movement. Our evaluation results show that \mech~\gcam{achieves} \hdiffcpu, \hdiffgpu, and \hdifffpga higher performance compared to the state-of-the-art CPU, GPU, and FPGA implementations \gom{of horizontal diffusion}, respectively. \gom{Compared to the  most  energy-efficient design on an HBM-based FPGA, \mech provides 2.43$\times$ higher energy efficiency.} Our results \gom{indicate} that   spatial architectures can mitigate current technological limitations for weather stencil computation. 

This work makes the following major contributions:

\begin{itemize}[leftmargin=*, noitemsep, topsep=0pt]
\item We perform a detailed analysis to show \gom{that} \gcam{the fundamental} weather prediction kernel \gcam{is constrained by}  memory bandwidth on state-of-the-art CPU, GPU, and FPGA-based systems. 

\item We introduce \mech, the first spatial accelerator 
for  horizontal diffusion stencil from a real-world  weather model. 
\gom{This work is} the first to propose and evaluate a real spatial architecture \gomm{specifically} for weather stencil computation\gom{, as opposed to commonly used machine learning-based workloads.}

\item We provide the first scaled-out spatial accelerator design using \gcam{the} multi-level intermediate representation (MLIR) framework. We analyze: 1) floating and fixed pipeline datapath\gom{s}, and 2) mu\gom{l}ti-core design using different interconnects on a real cutting-edge spatial architecture, \gom{the} AMD-Xilinx Versal AI Engine.


\item We conduct an in-depth evaluation of \mech on a real system showing that it outperforms state-of-the-art CPU, GPU, and FPGA-based implementations in terms of execution time. We open-source our implementations, \gcam{including five elementary stencils that are commonly used as benchmark\gom{s} for stencil computation research,} to aid further research in accelerating stencil computation on spatial architectures at \MYhref[cyan]{https://github.com/CMU-SAFARI/SPARTA}{https://github.com/CMU-SAFARI/SPARTA}. 

\end{itemize}


%% file: sections/02-background.tex
\section{Background}

\subsection{\gcam{Horizontal Diffusion Weather Stencil}}
\label{subsec:back:cosmo}
A stencil operation sweeps over an input grid, updating values based \gom{using a} fixed  \gom{computation} pattern. High-order stencils are applied to multidimensional grids that have sparse and irregular memory access patterns, limiting the achievable performance. In addition, stencils have  limited cache data reuse which further enhances memory access pressure.     
Unlike stencils found in the literature~\cite{waidyasooriya2019multi,singh2019low,de2021stencilflow,sano2014multi,7582502,chi2018soda,de2018designing}, real-world compound stencils\gom{~\cite{gysi2015modesto,cosmo_knl}} consist of a collection of stencils that perform a sequence of element-wise computations with complex interdependencies. Horizontal diffusion (\hdiff) represents \gcam{one such fundamental compound stencil} 
found in the \gom{dynamical core}  of the COSMO \gom{weather} model\gom{~\cite{gysi2015modesto,cosmo_knl}}. 

Horizontal diffusion is a process used in numerical weather prediction models to help smooth out small-scale variations in the atmosphere and reduce the impact of numerical errors. It is a mathematical technique used to represent the turbulent mixing of the atmosphere, which occurs due to the winds and temperature gradients in the atmosphere. Horizontal diffusion is used in a variety of other local and global weather prediction models~\cite{gysi2015modesto,cosmo_knl,de2021stencilflow,palmer1990european,mcclung2016global,hurrell2013community,watanabe2011miroc}. We can write the \hdiff equations as follows~\cite{cosmo_knl}:

\begin{gather}
\scalebox{0.92}{$
L_{r,c,d}^{n}\!=4\psi_{r,c,d}^{n}\!- \psi_{r+1,c,d}^{n}\!-\psi_{r-1,c,d}^{n}\!-\psi_{r,c+1,d}^{n}\!-\psi_{r,c-1,d}^{n} \label{eq:lap}$}\\[-6pt]
\scalebox{0.92}{$F_{r+\frac{1}{2},c,d}^n\!$}=
    \begin{cases}
    L_{r+1,c,d}^n\!-L_{r,c,d}^n\! & if (L_{r+1,c,d}^n\!-L_{r,c,d}^n)\\&(\psi_{r+1,c,d}^n\!- \psi_{r,c,d}^n) \leq 0\\
    0 & otherwise
    \end{cases} \label{eq:f}
\end{gather}
\begin{gather}
\scalebox{0.92}{$G_{r,c+\frac{1}{2},d}^n\!$}=
    \begin{cases}
    L_{r,c+1,d}^n\!-L_{r,c,d}^n\! & if (L_{r,c+1,d}^n\!-L_{r,c,d}^n)\\&(\psi_{r,c+1,d}^n\!- \psi_{r,c,d}^n) \leq 0\\
    0 & otherwise
    \end{cases} \label{eq:g}\\[-6pt]
   \scalebox{0.92}{$ \Psi_{r,c,d}^{n+1}=\psi_{r,c,d}^{n}\!-C_{r,c,d}^{n}\!(F_{r+\frac{1}{2},c,d}^n\!-F_{r-\frac{1}{2},c,d}^n\!$}\label{eq:out}\\\notag \scalebox{0.92}{$+G_{r,c+\frac{1}{2},d}^n\!-G_{r,c-\frac{1}{2},d}^n\!) $}
\end{gather}
\vspace{0.1cm}

$\psi$ represents the input data field at timestep $n$. $L_{r,c,d}^{n}$ is the discretized Laplacian \gom{(or Laplace operator)} in 
Equation~\ref{eq:lap} over row, column, and depth dimensions of an input grid. While  $F_{r+\frac{1}{2},c,d}^n$ and $G_{r,c+\frac{1}{2},d}^n$ represent limited fluxes along horizontal axes in 
Equations~\ref{eq:f} and \ref{eq:g}, respectively. In 
Equation~\ref{eq:out}, $\Psi_{r,c,d}^{n+1}$ represents the diffused output using a diffusion coefficient, $C_{r,c,d}^{n}$. The horizontal diffusion coefficients are used to represent the strength of the mixing, and are typically based on the physical properties of the atmosphere, such as  wind speed and temperature. The coefficients are usually chosen based on the model resolution, and can be adjusted to account for the effects of small-scale atmospheric processes.




\hdiff iterates over a 3D grid performing \textit{Laplacian} and \textit{flux}, as depicted in Figure~\ref{fig:hdiff}, to calculate different grid points. \gom{Each color represents a different stencil pattern that is a part of the hdiff computation.}  A \textit{Laplacian} stencil accesses the input grid at five memory offsets in horizontal dimensions. The \gom{output of} \gommm{the} \textit{Laplacian} \gom{stencil is used to calculate \gommm{the}  \textit{flux} stencil. The \textit{flux} stencil also requires access to the} input data for computation \gom{(indicated by a dashed arrow in Figure~\ref{fig:hdiff})}.   \hdiff has data dependencies in the horizontal neighborhood, therefore, we \gs{can} parallelize \hdiff in the vertical dimension. 

\input{images/hdiff}

Algorithm~\ref{algo:hdiffKernel} shows a pseudo-code for the fourth-order \hdiff~\gom{(i.e., the order of the highest derivative, also known as differential coefficient, is four for \hdiff).}

\input{algorithms/cosmo}

Figure~\ref{fig:hdiffMemory} shows the memory layout for the horizontal diffusion kernel. \hdiff~\gom{computation} typically involves a large amount of data \gom{access and} manipulation. We observe that \gom{\hdiff has complex input data} access \gom{patterns that} can impact  cache efficiency by incurring a significant \gomm{number} of cache misses, leading to sub-optimal 
performance on our current systems. The data that is used for \hdiff is typically stored in a multi-dimensional array \gom{(referred to as a \emph{grid})}. \gom{\hdiff computation requires access to input data elements in memory that are located both close to each other in memory (i.e., local memory access) and far apart in memory (i.e., global memory access).} The local memory access patterns involve accessing the  \gom{neighboring grid cells within the same row or column. Since the input data is stored contiguously in memory, local access is relatively fast due to the spatial locality for which cache lines can be efficiently utilized.} 
However, global memory access patterns \gom{\gomm{require} grid cells that are distant from the current cell in the input grid. Since the input data is not stored contiguously, global memory access is generally slower than local memory access. This type of access can lead to increased cache misses and latency. Therefore, it is crucial to optimize the memory access patterns to improve the performance of the \hdiff computation.} 
\input{images/hdiff_layout}

\subsection{Spatial Computing Architectures}
\label{subsec:back:aie}

\begin{figure}[h]
\centering
\vspace{0.25cm}
\includegraphics[width=0.45\textwidth]{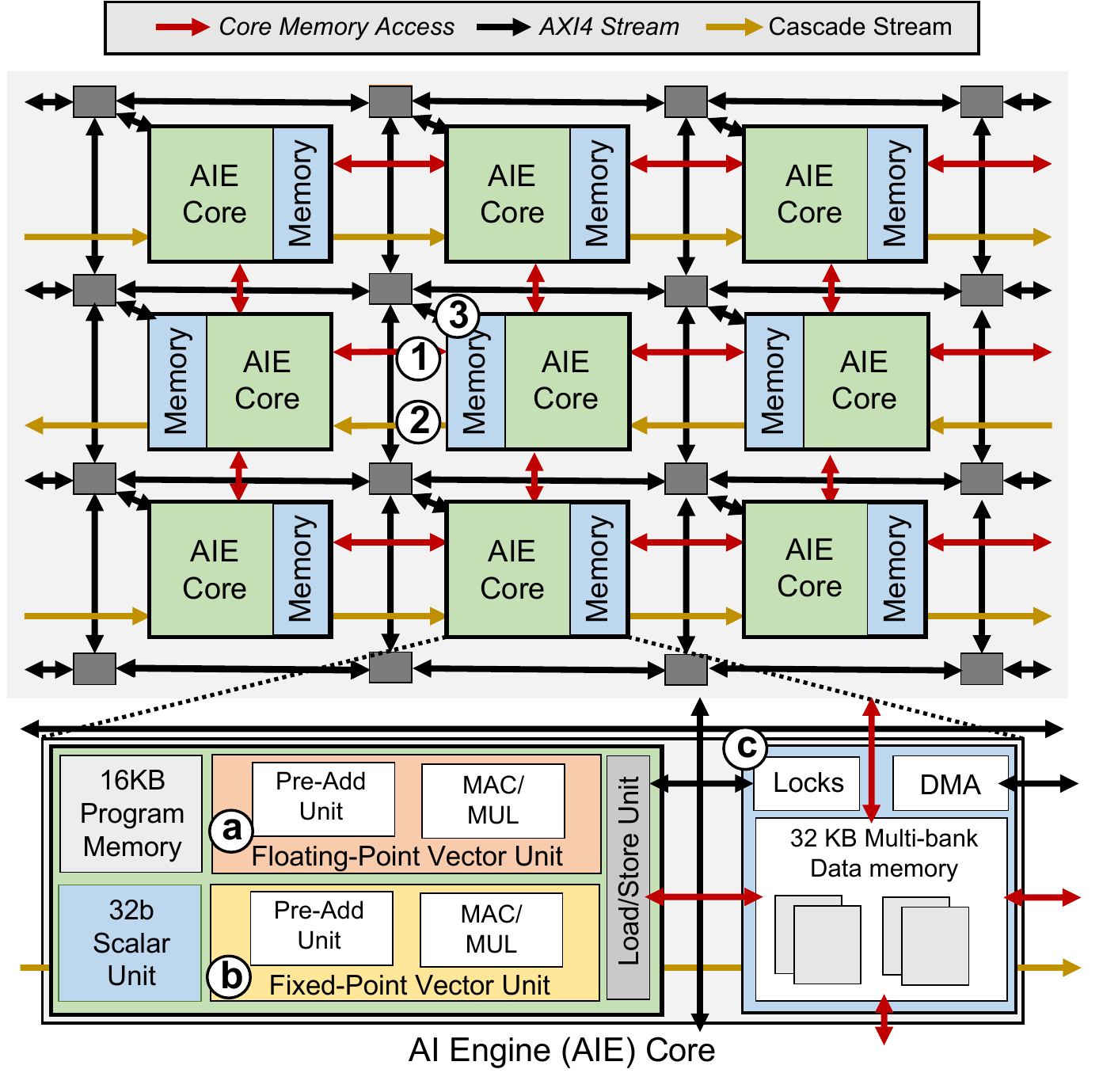}
\vspace{-0.15cm}
\caption{AI Engine (AIE) architecture overview. AIE consists of an array of processing cores interconnected through a highly configurable network. Each AIE core has independent datapaths for fixed-point and floating-point instructions with 32KB of local \gom{data} memory.}
\label{fig:aie_array}
\end{figure}

A spatial architecture \gom{(}such as~\cite{abts2020think,la2020cerebras,morgan2018intel,vissers2019versal}\gom{)} consists of two-dimensional interconnected processing units.  In AMD-Xilinx Versal~\cite{vissers2019versal},  such a unit is referred to as an AI Engine (AIE) core\gommm{,} as shown in Figure \ref{fig:aie_array}. 

The Versal platform has 400 of such \aie cores arranged in a checkerboard fashion. Each \aie~\gom{core} ~\gcam{contains} 
its own \gom{local} program memory (16KB) and \gom{local} data memory (32KB) \gom{and} can execute instructions independently of all other cores.   The \aie~\gom{core}  is a VLIW processor that can issue up to seven instructions per cycle divided over two datapaths: a scalar and a vector. Due to these two independent datapaths, \gom{an}~\aie~\gom{core}  can perform complex pointer arithmetic using the scalar datapath \gom{while} fully utilizing the vector datapath. Each \aie~\gom{core}  is clocked at 1GHz and programmed through C/C++ code using \aie intrinsics~\cite{AIEintrinsic}. The vector datapath implements two-dimensional SIMD operations for both floating-point \circledWhite{a} and fixed-point vector units \circledWhite{b}. The fixed-point datapath supports precisions ranging from \gcam{8-bit} to 32-bit 
operands, which corresponds to 128 \gom{multiply and accumulate operations (MACs)} to 8 MACs per cycle. \gom{The floating-point and fixed-point datapaths provide a \textit{pre-add}
unit for doing vector elementary functions, such as determining the
minimum or maximum of two vectors or comparing two vectors.} \gom{The AIE local data memory module \circledWhite{c} is divided into eight
memory banks and also consists of a direct memory access (DMA) interface and locks. Locks allow synchronization between
AIE cores, an AIE core and DMA, and an external memory-mapped AXI4 device (outside
of the AIE array).} 



An \aie~\gom{core} can communicate with other \aie cores using three different interfaces~\cite{AIEngineArch}\gom{, as shown in Figure~\ref{fig:aie_array}}: \circledWhite{{1}} \emph{Core memory access}: direct load/store access to the memories of immediate neighboring AIE cores  \gom{via} a 256-bit interface,  \circledWhite{{2}} \emph{Cascade stream}: enables the transfer of (full accumulator precision) partial sums to a direct neighbor \gom{via} a 384-bit interface, and \circledWhite{{3}} \emph{AXI4 stream}:  enables \gom{DMA}  
to non-immediate AIE \gom{cores} via two AXI-4 streams \gom{(also called as stream switches)} of 32-bit.   The core memory  access interface allows access to \gommm{the} memory \gom{of neighboring AIE cores}, leading to a total of 128-KiB contiguous \gom{data} memory space \gom{(i.e., its own memory, the memory of the AIE core on the north, the  memory of the AIE core on the south, and the  memory of the AIE core on the east or west depending on the row and the relative placement of \aie core and memory module)}. 
The AXI4 stream interface can be configured to duplicate data to different AIE cores 
to enable input data \emph{broadcast}\gom{, which allows multiple AIE cores to share the same input data.} 
AIE also consists of 16 dedicated \textit{shimDMA} (shim Direct Memory Access) cores in the bottom row for interfacing with external dynamic random-access (DDR) memory. 
\gom{A shimDMA is a lightweight DMA engine that provides a simplified interface for data movement between system memory and hardware accelerators. AIE consists of two 256-bit channels per shimDMA for read and write operations, respectively.} 




%% file: images/hdiff.tex
\begin{figure}[H]
\centering
\resizebox{0.39\textwidth}{!}{
  \begin{tikzpicture}[rotate=90,transform shape]
     \gridThreeD{0}{8.25}{black};
    \node at (7.58,10.5)[ rotate=270] {\Large \textit{Laplacian}};
    \node at (7.2,10.5)[ rotate=270] {\Large\textit{Stencil}};
    \gridThreeDSecond{1}{4}{black};
    \node  at (7.58,5.5)[ rotate=270] {\Large \textit{Flux}};
    \node at (7.2,5.5)[ rotate=270] {\Large \textit{Stencil}};
    \gridThreeDThird{2}{0}{black};
     \node  at (7.58,1) [ rotate=270] {\Large \textit{Output}};
      \draw [thick,->,blue] (2.75,4.25) -- (2.7,8.4);
      \draw [thick,->,blue] (2.75,4.25) -- (2.1,8.9);
      \draw [thick,->,blue] (2.75,4.25) -- (4.1,8.9);
      \draw [thick,->,blue] (2.75,4.25) -- (3.5,9.4);
      \draw [thick,->,orange] (2.1,4.7) -- (3.5,9.4);
      \draw [thick,->,orange] (2.1,4.7) -- (1.5,9.4);
      \draw [thick,->,orange] (2.1,4.7) -- (2.1,8.9);
      \draw [thick,->,orange] (2.1,4.7) -- (2.9,9.9);
      \draw [thick,->,red] (4.1,4.7) -- (3.5,9.4);
      \draw [thick,->,red] (4.1,4.7) -- (5.5,9.4);
      \draw [thick,->,red] (4.1,4.7) -- (4.1,8.9);
      \draw [thick,->,red] (4.1,4.7) -- (4.9,9.9);
          \draw [thick,->,brown] (3.5,5.25) -- (3.5,9.4);
          \draw [thick,->,brown] (3.5,5.25) -- (2.9,9.9);
          \draw [thick,->,brown] (3.5,5.25) -- (4.9,9.9);
          \draw [thick,->,brown] (3.5,5.25) -- (4.3,10.4);
         \draw [thick,->,violet] (2.7,0.25) -- (3.5,5.15);
         \draw [thick,->,violet] (2.7,0.25) -- (4.1,4.65);
         \draw [thick,->,violet] (2.7,0.25) -- (2.1,4.65);
         \draw [thick,->,violet] (2.7,0.25) -- (2.75,4.15);
    \draw [thick,->,violet, dashed] (4.1,8.9) .. controls (8.1,8.7) and (3.7,0.25) ..  (2.7,0.25);
\end{tikzpicture}
}
\caption{\hdiff kernel composition {using Laplacian and flux stencils} in a two\gommm{-}dimensional plane~\cite{narmada}. \gom{Each color represents a different stencil pattern that is a part of the \hdiff computation. The dashed arrow represents the dependency of \gommm{the}~\flx stencil on the input data.}\label{fig:hdiff} }
\end{figure}

%% file: algorithms/cosmo.tex
\begin{algorithm}[h]
\footnotesize
\SetAlgoLined
\DontPrintSemicolon
\SetNoFillComment
\caption{Pseudo-code for horizontal diffusion (\hdiff) {kernel}
used by the COSMO~\cite{doms1999nonhydrostatic} {weather prediction} model.}
\label{algo:hdiffKernel}
\SetKwFunction{FMain}{hdiff}
\SetKwProg{Fn}{Func}{}{end}
  \Fn{\FMain{float* src, float* dst}}{
       \For{$d\gets1$ \KwTo $depth$}{
             \For{$c\gets2$ \KwTo $column-2$}{
                    \For{$r\gets2$ \KwTo row-2}{
                        \tcp*[l]{{L}aplacian calculat{ion}}
                        $lap_{CR}=laplaceCalculate(c,r)$ \label{algo:line:lap_cr}
                        
                        \tcp*[l]{row-laplacian}
                        $lap_{CRm}=laplaceCalculate(c,r-1)$\;   \label{algo:line:lap_crm}   
                        $lap_{CRp}=laplaceCalculate(c,r+1)$
                        \tcp*{column-laplacian}
                        $lap_{CmR}=laplaceCalculate(c-1,r)$\;
                        $lap_{CpR}=laplaceCalculate(c+1,r)$
                        \tcp*{column-flux calculat{ion}}\
                        $flux_{C} = lap_{CpR} - lap_{CR}$\;
                        $flux_{Cm} = lap_{CR} - lap_{CmR}$\;
            \tcp*[l]{row-flux calculat{ion}}
                        $flux_{R} = lap_{CRp} - lap_{CR}$\;
                        $flux_{Rm} = lap_{CR} - lap_{CmR}$\;
                        \tcp*[l]{output calculat{ion}}
                        $dest[d][c][r] = src[d][c][r] -
                            c1 * (flux_{CR}- flux_{CmR}) + (flux_{CR}- flux_{CRm})$
                        }
            
                }
   
         }
}
\end{algorithm}

%% file: images/hdiff_layout.tex
\begin{figure}[h]
  \resizebox{0.47\textwidth}{!}{
    \begin{tikzpicture}[scale=0.3]
  \inputLayout{}
    \laplines
    \fluxlines
    \outlines
  \end{tikzpicture}
     }
     \vspace{0.1cm}
\caption{Memory layout of horizontal diffusion from 3D input grid onto 1D memory array. \gom{Each color represents a different stencil pattern that is a part of the \hdiff computation. The dashed arrows represent the dependency of \flx stencil on the input data.} \label{fig:hdiffMemory}}

 \end{figure}

%% file: sections/03-implementation.tex
\section{Implementation}
\label{sec:implementation}
\gs{To design \mech, we analytically examine the comput\gommm{ation} and memory requirements of \hdiff for \aie (Section~\ref{subsect:implemetation_analytic}). Based on our analytic\gom{al} modeling, we \gom{propose,} \gommm{design, and implement} two different \aie-based architectures: (1) \emph{\saie}: single AIE core that performs both \lap and \flx calculations (Section~\ref{subsec:implementation_saie}), and (2) \emph{\maie}: multiple AIE cores that compute \lap and \flx on separate \aie cores (Section~\ref{subsubsec:implementation_maie}). }


\subsection{Analytical Modeling}
\label{subsect:implemetation_analytic}

\head{\gom{Analysis of} Comput\gom{ation} Requirements} We first calculate the comput\gom{ation} cycles requirement for \hdiff.
\hdiff has five \lap stencils followed by four \flx stencils, \gcam{as shown in Algorithm~\ref{algo:hdiffKernel}}. In every iteration, \hdiff requires  access to data from five rows of the input grid. Each \lap stencil performs five MAC operations, which are applied throughout the input grid except for the border region leading to $5 \times (R-4)\times (C-4)$ MAC operations, where R and C represent the number of rows and columns of a grid, respectively.  As these operations are applied throughout the vertical dimension for all five \lap stencils, the total number of MAC operations is $5 \times (R-4)\times (C-4)\times D\times 5$, where D represents the number of planes of the input grid. 

Each \flx stencil requires 2 MAC operations, 1 subtract operation, 1 compare operation, and 1 select operation. In total, for the four \flx stencils, we require $2 \times (R-4)\times (C-4)\times 4$ MAC operations and $1 \times (R-4)\times (C-4)\times 4$ operations each for subtract, compare, and select operations. The subtract, compare, and select operations use only the pre-add units in the vector datapath and not the MAC units.  

In this work, we use the int32 and float32 data format due to the high-precision requirements in weather modeling~\cite{bauer2021digital}. \gcam{\gom{int32 and float32} formats are widely used for weather modeling, depending on the specific requirements~\cite{vavna2017single,kimpson2023climate,palmer2019stochastic,hatfield2019accelerating,chantry2021opportunities,chantry2021machine,saffin2020reduced,yuval2021use,klower2020number,paxton2022climate,ackmann2022mixed}, e.g., regional weather modeling, ensemble modeling, and short-term weather forecasting. }  
The use of higher precision allows for: (1) more accurate modeling of the complex interactions between the various physical processes involved in weather, such as temperature, pressure, moisture, and wind, and (2) the simulation of a wide range of time scales, from small-scale processes like cloud formation to large-scale phenomena, such as global weather patterns. Therefore, we can have a more accurate representation of the dynamics of weather systems over time. As a single \aie core is capable of 8$\times$
32-bit multiply-and-accumulate (MAC) operations in one clock cycle, the minimum number of cycles required for a complete sweep of \hdiff on the input grid can be calculated using: 
\begin{gather}
\small
 \lap_{comp}= \frac{5 \times (R-4)\times (C-4)\times D\times 5}{8}\label{comp_lap}\\
 \flx_{comp}= \frac{2 \times (R-4)\times (C-4)\times D\times 4}{8}\label{comp_flx}\\\notag
                + \frac{3\times (1 \times (R-4)\times (C-4)\times D\times 4)}{8}\\
\hdiff_{comp}=\lap_{comp}+\flx_{comp}\label{comp_hdiff}
\end{gather}
\vspace{0.01cm}

Where R, C, and D represent the number of rows, columns, and planes, respectively, of the input grid. This calculation assumes that all the MAC operations can be performed in parallel \gom{with 100\% MAC efficiency.} 

\head{\gom{Analysis of} Memory Requirements}
Next, we 
analyze the memory intensity \gs{(memory cycles)}
of \hdiff. The coefficients for \lap and \flx stencils can be stored in the vector registers
, \gom{which avoids} the need to  
fetch \gom{the coefficients} from an \aie~\gom{core} data memory or shimDMA for each MAC operation. However, the input grid data needs to be tiled and loaded into the local data memory from the external memory for the MAC operations. 
An \aie~\gom{core} supports two 256-bit loads per cycle using the core memory access interface. \lap needs to access $5 \times (R-4)\times (C-4)\times D\times 5$ elements and \flx needs $2 \times (R-4)\times (C-4)\times D\times 4$ elements for a complete \hdiff calculations. For 32-bit precision, the minimum number of memory cycles required for a complete sweep of hdiff on the input grid can be calculated as:  

\begin{gather}
\small
 \lap_{mem}= \frac{5 \times (R-4)\times (C-4)\times D\times 5\times 32}{2\times256}\label{mem_lap}\\
 \flx_{mem}= \frac{2 \times (R-4)\times (C-4)\times D\times 4\times 32}{2\times256}\label{mem_flx}\\
 \hdiff_{mem}=\lap_{mem}+\flx_{mem}\label{mem_hdiff}
\end{gather}
\vspace{0.01cm}

\textbf{Discussion.}
\gcam{\gcam{The above} high-level modeling provides a first-order estimate of external memory traffic and comput\gommm{ation} requirements,  enabling the identification of parallelization and data-reuse opportunities.} We make the following two observations. First, the \lap stencils have a more balanced design than \flx stencils because the comput\gommm{ation}-to-memory intensity  ratio of the \lap stencils (as represented by Equation~\ref{comp_lap} and Equation~\ref{mem_lap}, respectively) is more balanced 
 compared to the comput\gommm{ation}-to-memory intensity ratio of  the \flx stencils(as represented by Equation~\ref{comp_flx} and Equation~\ref{mem_flx}, respectively). 
The \flx stencils have a higher compute-bound (as represented by Equation~\ref{comp_flx}) than the memory bound (as represented by Equation~\ref{mem_flx}). Second, the non-MAC operations (subtract, compare, and select) in the \flx stencil lead to higher comput\gom{ation} cycles due to the frequent movement of data between accumulator and vector registers.

Our analytical modeling of \hdiff on AIE leads to the insight that achieving the maximum throughput requires splitting the hdiff computation over multiple AIE cores. 
By splitting the \hdiff computation over multiple cores, we \gom{obtain} two main benefits. First, the compute-bound can be distributed among multiple cores, allowing for the concurrent execution of multiple stencil calculations, which can increase the overall performance and throughput of the \hdiff algorithm. Second, it allows for the use of more AIE cores in parallel per shimDMA to achieve higher throughput. 

\subsection{Mapping onto  the \aie Cores}

For both our designs \gom{(\saie and \maie)}, we carefully hand-tune the \hdiff code to overlap memory operations with arithmetic operations, to improve performance. In \maie, we use the AIE data forwarding interfaces to forward the results from the first AIE core (used for \lap calculation) to another AIE core (used for \flx calculation). This approach allows for the concurrent execution of multiple stencil calculations \gcam{to} increase the overall performance and throughput of the \hdiff design.

\subsubsection{Single-AIE Core Mapping} 
\label{subsec:implementation_saie}
\begin{figure}[h]
  \centering
  \includegraphics[width=\linewidth,trim={0cm 0cm 2.5cm 0cm},clip]{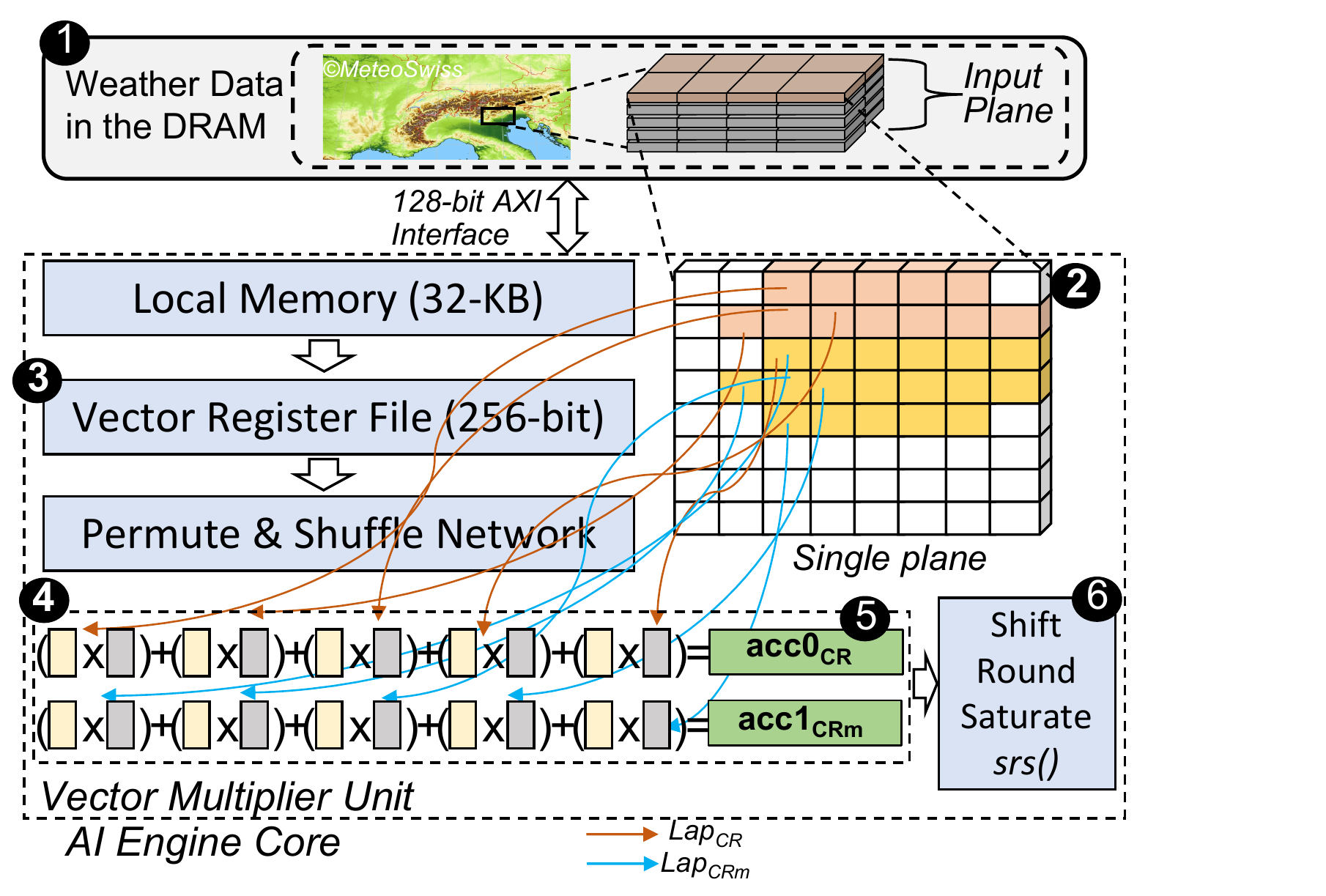}
  \caption{Mapping of two \lap stencils onto an \aie core datapath. We use the same AIE core to perform the \lap and \flx computations. The weather data from the host DRAM is buffered into \gom{the} local memory of an AIE core \gcam{and further loaded} into vector registers to perform \hdiff computation. \label{fig:aie_hdiff_lap}}
 \end{figure}

\fig{\ref{fig:aie_hdiff_lap}} shows the  mapping of two \lap stencils, $lap_{CR}$ and $lap_{CRm}$ from Algorithm~\ref{algo:hdiffKernel} (out of the five \lap stencils and four \flx stencils needed for a single \hdiff computation) onto an AIE core. We use the same AIE core to perform the remaining \lap and \flx computations. 

The weather data, based on the atmospheric model resolution grid, is stored in the DRAM of  host system \circled{1}. We parallelize the computation across input planes. \gcam{The limited on-chip memory per AIE core makes buffering entire 3D planes \gom{in}feasible. Therefore, we tile the plane into smaller blocks with a sliding window that can fit into the on-chip memory.} A single \hdiff output computation requires access to data from five input rows of a single plane \circled{2}. Therefore, five rows are loaded into the local memory of an AIE core using a shimDMA core. We employ the double buffering (ping-pong) technique between the shimDMA and the core local memory to hide the transfer latency. From the local memory \circled{3}, the data is loaded into vector registers to perform the computations required for all the different \lap and \flx stencils (see Section~\ref{subsec:back:cosmo}). 
We vectorize all the operations, e.g., $lap_{CR}$ and $lap_{CRm}$ perform five element-wise operations, each using the vector multiplier unit \circled{4}. We store partial results from each vector operation of a stencil in the same accumulator register \circled{5}. 
The programmer must ensure that there are not too many live registers (both vector and accumulator registers) as this can cause register spilling (i.e., the compiler would spill data onto the stack memory), resulting in wasted clock cycles.
We allocate vector and accumulator registers among \lap and \flx stencil computation making sure that there is no unnecessary spilling.
It is essential to perform as many operations as possible while the data is in the accumulator register. We observe that the $lap_{CR}$ calculation, as described in Algorithm~\ref{algo:hdiffKernel}, is utilized for all four subsequent \flx stencil computations. Therefore, to minimize computation cycles and re-computation of results, we: (1) store the $lap_{CR}$ accumulator
  results in a dedicated vector register using the shift-round-saturate intrinsic (\srs)
  , and (2) reuse the $lap_{CR}$ accumulator to perform as many flux stencil operations as possible. 
 
 
 For \flx stencil, the non-MAC operations, such as vector addition/subtraction, vector compare, and vector selection, do not store data in an accumulator register as it uses \gcam{the} pre-adder unit that only has access to vector registers. Therefore, to continue with \flx calculations after performing \lap operations, we need to send the data back from the accumulator registers to the vector registers using \gom{an}~\srs operation (from shift round saturate unit \circled{6} to vector register file \circled{3}). 
 This operation has a long latency (4 cycles), which reduces the number of instructions that can be completed within one cycle. Therefore, this process of moving data between registers can negatively impact performance and requires manual hand-tuning of the algorithm to hide data transfer latency. To overcome this, we carefully rearrange load/store, MAC/MUL, and non-MAC operations to fill the VLIW pipeline of the AIE core while avoiding NOPs. This rearrangement assists the compiler in proper scheduling and prevents multiple sequential load operations \gom{from} fill\gom{ing} the vector registers, which could lead to \gom{wasted} VLIW \gom{instruction} slots and decreased performance. 
 
 \subsubsection{Multi-\aie Core Mapping}
 \label{subsubsec:implementation_maie}

\begin{figure}[h]
  \centering
\includegraphics[width=0.9\linewidth,trim={0.2cm 0cm 1cm 0cm},clip]{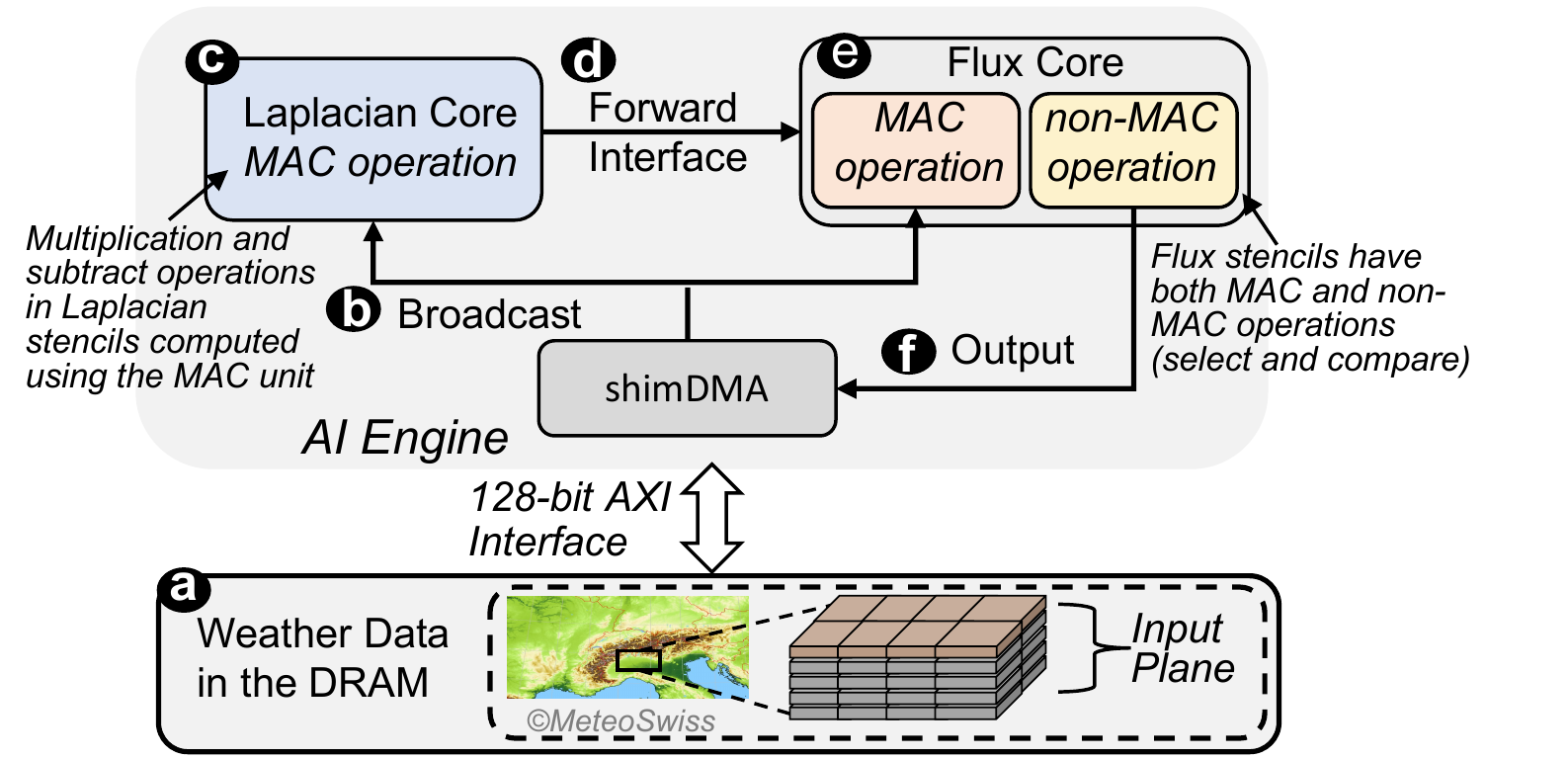}
  \caption{Multi-AIE design for \hdiff computation to balance the comput\gommm{ation} and the memory bound. We show the dataflow sequence from the DRAM memory to the AIE cores via shimDMA.\label{fig:multi_hdiff}}
 \end{figure}

 Based on our analytical modeling \gcam{(Section~\ref{subsect:implemetation_analytic})}, we \gom{create} two \maie designs: dual-AIE and tri-AIE.  \fig{\ref{fig:multi_hdiff}}  shows the multi-AIE design approach for \hdiff, where the \flx stencil uses the results of the \lap stencil to perform its computation. We also show the dataflow sequence from the external DRAM memory to the AIE cores. Instead of waiting for the Laplacian AIE core  \circled{c}  to complete all five Laplacian stencils required for a single hdiff output, we forward the result \gom{of} each Laplacian stencil to the flux AIE core  \circled{e}, thereby allowing both cores to remain active. In tri-AIE design, we further split \flx computation and map MAC operation\gom{s} and non-MAC operations onto different AIE cores.

As described in Algorithm~\ref{algo:hdiffKernel}, both \lap and \flx computations require access to the input data, which is stored in the external DRAM \circled{a}. Therefore, we broadcast \circled{b} the input data onto the local memor\gom{ies} of both Laplacian and Flux AIE cores using a single shimDMA channel. In \daie design, a single Flux core is responsible for performing all MAC and non-MAC operations.  As discussed in Section~\ref{subsect:implemetation_analytic}, \flx operations have an imbalance between comput\gommm{ation} and memory bounds. Therefore, to further improve the compute performance, we split \flx operations over two AIE cores in our \taie design. We utilize the data forwarding interfaces to forward  \circled{d}  the results of the \lap stencil computations from the first AIE core to the subsequent AIE core for \flx stencil computation. 
After computing the output results, we transfer the results from the local memory of an AIE core to the external DDR memory  \circled{f}.

\subsection{Managing Data Transfer using MLIR}
\gcam{Versal provides a large number of AIE cores that can be interconnected using a wide range of interfaces. This complicates the effective management of dataflow between all the available cores.} We utilize MLIR (Multi-Level Intermediate Representation)~\cite{mlir} as a way to separate the AIE core computation optimization and code generation process in a flexible and modular manner. Our approach involves hand-optimizing code for an AIE core datapath and using MLIR to generate low-level code to connect the AIE cores and manage data transfer between the external DDR memory and AIE core memory. 
 
 We provide a code \gom{example} for our \saie design in Code~\ref{lst:hdiffmlir}. We instantiate an AIE core and a shimDMA core (line\gom{s}~\ref{line:aiecore}-\ref{line:shimdma}) \gcam{to perform \hdiff computation and transfer data from  external memory to AIE core memory, respectively}. 
 We create two AIE object FIFOs for input and output data (line\gom{s}~\ref{line:fifoin}-\ref{line:fifoout}).  These FIFOs are circular queues~\cite{denolf2007exploiting} that buffer input data to exploit the data reuse in \hdiff. We  also create two external memory buffers to read (write) data to (from) the external DDR memory (line\gom{s}~\ref{line:extin}-\ref{line:extout}). The AIE core 
 inside a for-loop (line~\ref{line:core_loop}), acquires five elements from the input object FIFO for consumption (line~\ref{line:core_acquire_in_sub}) and one element from the output object FIFO for production (line~\ref{line:core_acquire_out_sub}) as \hdiff requires five input grid rows to calculate a single output row. The \hdiff computation is performed by calling a function @vec\_hdiff (line~\ref{line:core_func_call}) with the acquired subviews of input and output FIFO\gom{s} \gcam{(line\gom{s}~\ref{line:core_acquire_in_sub}-\ref{line:core_acquire_out_sub})}. In every loop iteration, we release one input row from the input FIFO to advance to the next row in the input grid \gcam{(line~\ref{line:core_release_in_sub}) while producing an output row (line~\ref{line:core_release_out_sub})}. After \hdiff computation, all the remaining elements acquired from input FIFO are released \gcam{(line~\ref{line:core_release_in_remaing_sub})}, and the final results are transferred to the output DDR buffer \gcam{using a shimDMA write channel}.
\input{algorithms/mlir.tex}
 
\subsection{Scaling Accelerator Design}
The performance of an implementation can be maximized by scaling it out across as many AIE cores as possible while avoiding an imbalance in processing resources caused by data starvation (i.e., ensuring that each core has access to the data it needs to complete its task). However, scaling a design on all the processing cores available on spatial computing systems is a non-trivial task. As discussed in Section~\ref{subsec:back:aie}, there are only 16 shimDMA cores for 400 AIE cores. System architects need to develop a design that can balance comput\gommm{ation}, memory,  and \gcam{communication} resources. We observe four key challenges to scaling on spatial architectures: (1) balancing comput\gommm{ation} and memory resources, (2) limited shimDMA channels, (3) gathering and ordering of \gcam{output results before sending \gommm{them} back to the external memory}, and (4) placing input and output cores close to shimDMA to optimize data transfer. These challenges must be addressed in order to effectively scale the \hdiff implementation across multiple AIE cores.

\begin{figure}[h]
  \centering
\includegraphics[width=\linewidth,trim={0.3cm 0cm 0.5cm 0cm},clip]{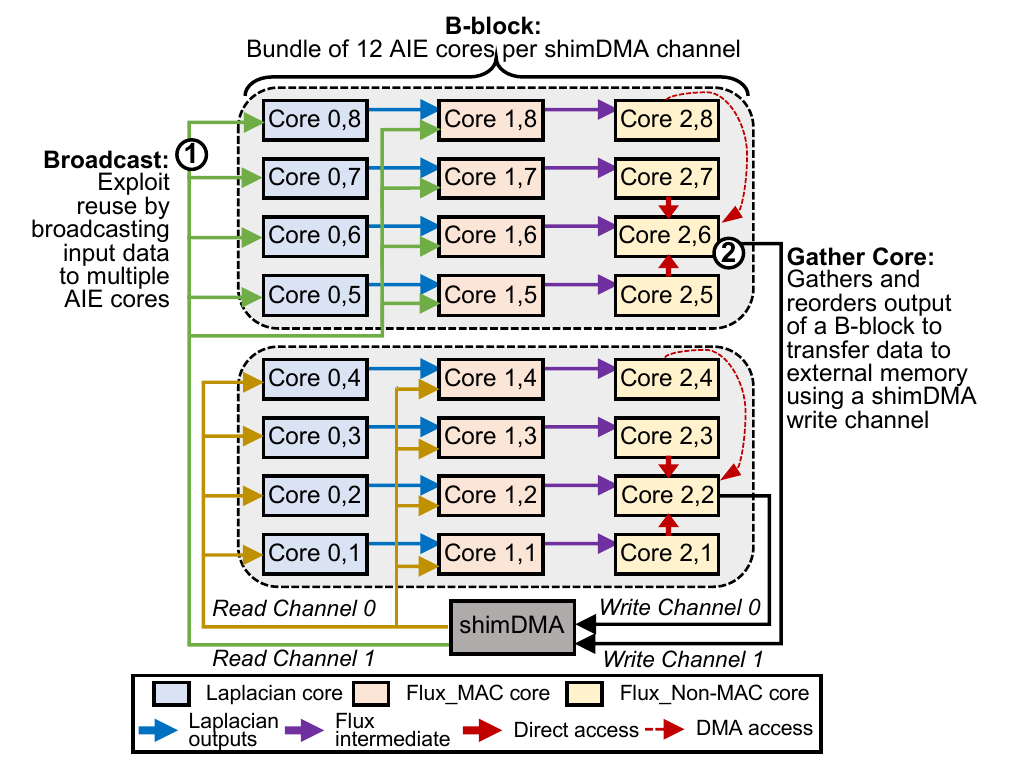}
  \caption{Block-based design (B-block) using \taie implementation for \hdiff. The \gom{B}-block design allows scaling computation across the AIE cores while balancing comput\gommm{ation} and communication time without getting bottlenecked by limited shimDMA channels.\label{fig:multiCoreHDIFF}}
 \end{figure}

We address 
these challenges by developing an architecture that combines fewer off-chip data accesses with higher throughput for the loaded data. 
To this end, our accelerator design takes a data-centric approach~\cite{mutlu2019,mutlu2021primer_pim,ghose2019processing,teserract,singh2019near,singh2019napel,hsieh2016accelerating,7551394,ahn2015pim,googleWorkloads,singh2022accelerating,singh2018review,singh2022sibyl,vadivel2020tdo,corda2019platform,mutlu2021intelligent,boroumand2021google} that exploits dataflow spatial acceleration.
We propose a \textit{bundle} or \textit{B-block}-based design. A B-block is a cluster of AIE cores connected to the same shimDMA input/output channel. 
As shown in Figure~\ref{fig:multiCoreHDIFF}, clusters of AIE cores are connected to two channels of a shimDMA (one for input and one for output). Each B-block comprises of multiple \emph{lanes} (or rows) of our tri-AIE design, with each lane calculating a different offset of output result using a part of the input plane. Based on our extensive empirical analysis to balance comput\gommm{ation} and memory performance,  we choose four lanes for a B-block design.

 As each lane requires access to five rows of the input grid to perform a single \hdiff computation, we use the broadcast feature of the global interconnect to duplicate the eight rows of the input data into a circular buffer in the AIE cores of the first column \circledWhite{\textbf{1}}.    An 8-element circular buffer  allows all the cores in the B-block lanes to work on a different offset of the input data while having five input grid rows necessary to perform \hdiff computation.

\gcam{Many} spatial architectures lack support for automatically gathering and ordering computed outputs. To overcome this limitation, we use physical placement constraints to allow the AIE cores in the last column of a B-block to access a single shared memory of a dedicated AIE core, enabling data gathering. We refer to this core as the \textit{gather core} \circledWhite{\textbf{2}}. The gather core is responsible for collecting data from all other cores in a B-block, in addition to processing the results of its own lane. We use a core that lies in the middle of the B-block as the gather core and not  a core located at either the top or bottom of the B-block due to the limited number of DMAs available on an AIE core. \gom{Each AIE core has only two DMAs. By placing the gather core in the middle of the B-block, we can directly access the memory of neighboring cores, reducing the total number of DMA operations required. If the gather core was placed at the top or bottom lane of a B-block, additional DMA operations would be needed to access the memories of further away cores\gomm{,} leading to higher data movement overhead and reducing overall performance. Therefore, placing the gather core in the middle of the B-block helps to reduce the data movement overhead by minimizing the distance between the gather core and the other cores in the block.} 

A single B-block operates on a single plane of the input data.  Since two B-blocks can be connected to a single shimDMA, two planes can be served per shimDMA.   
This regular structure can then be repeated for all the \gcam{16} shimDMA \gcam{cores} present on \gcam{the AIE} device. \gom{Our B-block-based design is highly scalable and can be expanded to utilize 384 AIE cores  (16 shimDMAs $\times$ 2 B-blocks per shimDMA $\times$ 12 AIE cores per B-block) while maintaining efficient resource allocation. 
Therefore, our B-block approach  enables: (a) \textit{modularity:} simplifies scaling as a B-block can be easily replicated and expanded to form a bigger system that can solve larger problem sizes, (b) \textit{efficient resource utilization:}  ensures that every AIE core and shimDMA is utilized effectively, which prevents resource starvation and maximizes \gomm{system} throughput, and (c) \textit{load balancing:} ensures that all AIE cores in the B-block lanes are working on different offsets of the input data while having the necessary input grid rows to perform hdiff computation.
}
\gcam{
\subsection{\mech Application Toolflow}
Figure~\ref{fig:toolflow} shows the \mech application toolflow to support our architecture. We provide a B-block description to the \mech generator  \circledWhite{{a}}. The generator configures data movement across AIE cores, including appropriate input/output buffers and interconnect description. Our \mech generator is an elementary tool for generating stencil-based accelerators. It can be absorbed in the MLIR ecosystem as a \emph{dialect}  to create \gom{a} custom, high-level intermediate representation (IR) that is tailored to the stencil application domain. Our hand-tuned \mech kernel code is compiled using \gom{the} Vitis AIE Compiler~\cite{vitiscompiler} \circledWhite{{b}} to generate object files (\texttt{*.o}). These object files are linked in our MLIR dataflow description (\texttt{aie.mlir}) and passed through MLIR-AIE~\cite{xilinxMLIR} \circledWhite{{c}}. MLIR-AIE is a custom IR tailored to AIE that generates low-level executable files (\texttt{*.elf}) for a target AMD Versal device. A software-defined host API \circledWhite{{d}} handles offloading jobs to AIE cores with an interrupt-based queuing mechanism. This allows for minimal CPU usage (and, hence, power usage) during AIE operation.

  \begin{figure}[H]
  \centering
  \vspace{-0.1cm}
  \includegraphics[width=1\linewidth,trim={0cm 0cm 0cm 0cm},clip]{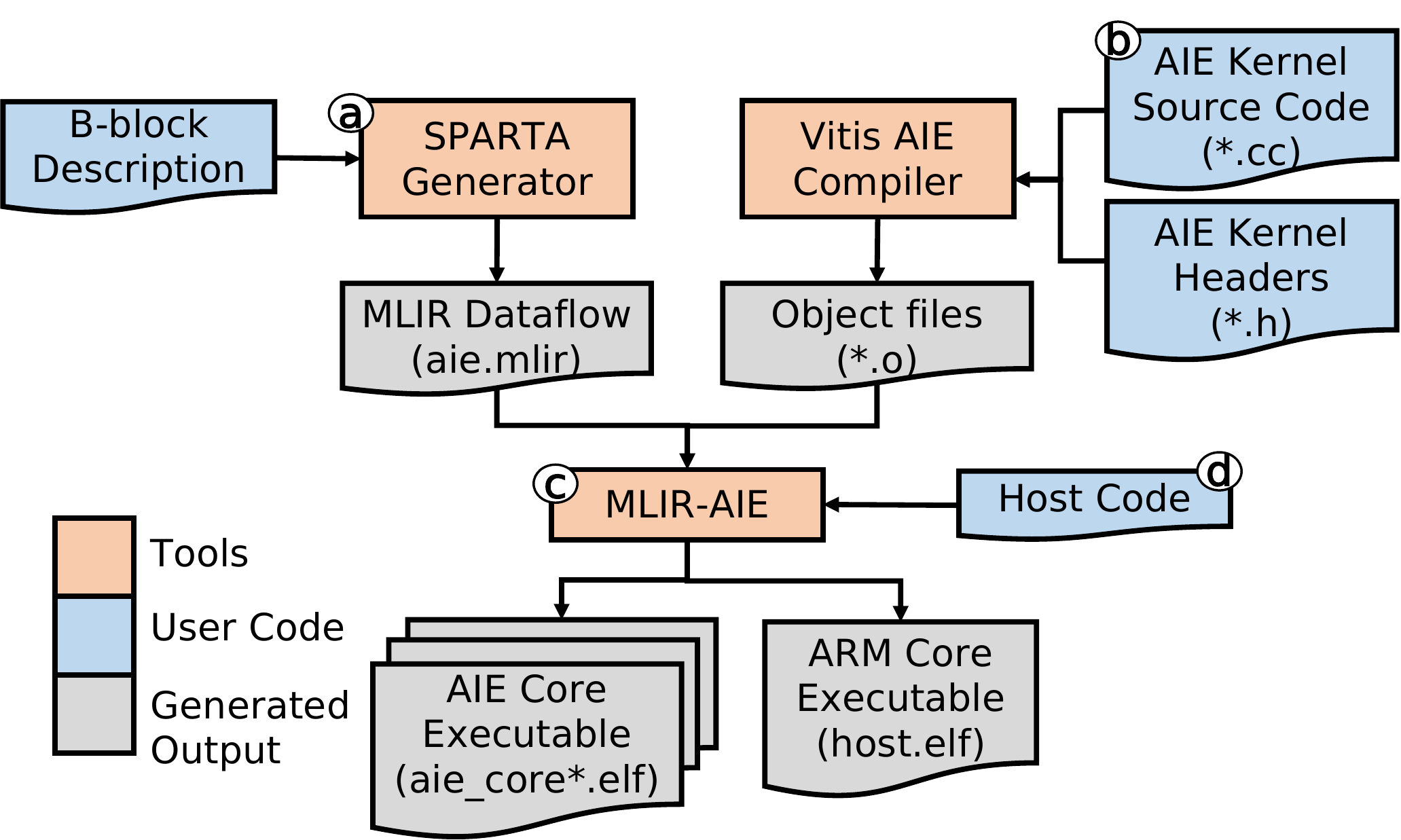}
  \caption{\mech application toolflow.\label{fig:toolflow}}
 \end{figure}
}

\subsection{Elementary Stencil Workloads}
We implement five elementary stencil workloads \gs{(\jacOneD\cite{pouchet2012polybench}, \jacTwoDT\cite{pouchet2012polybench}, \lap~\cite{doms1999nonhydrostatic}, \jacTwoDN\cite{pouchet2012polybench}, \sei\cite{pouchet2012polybench})} that are frequently used as benchmark\gom{s} for \gom{studying} stencil \gom{computations}~\cite{datta2008stencil,datta2009auto,datta2009optimization}. 

\begin{figure}[h]
  \centering
  \vspace{0.25cm}
  \includegraphics[width=1\linewidth,trim={0cm 0cm 2.6cm 0cm},clip]{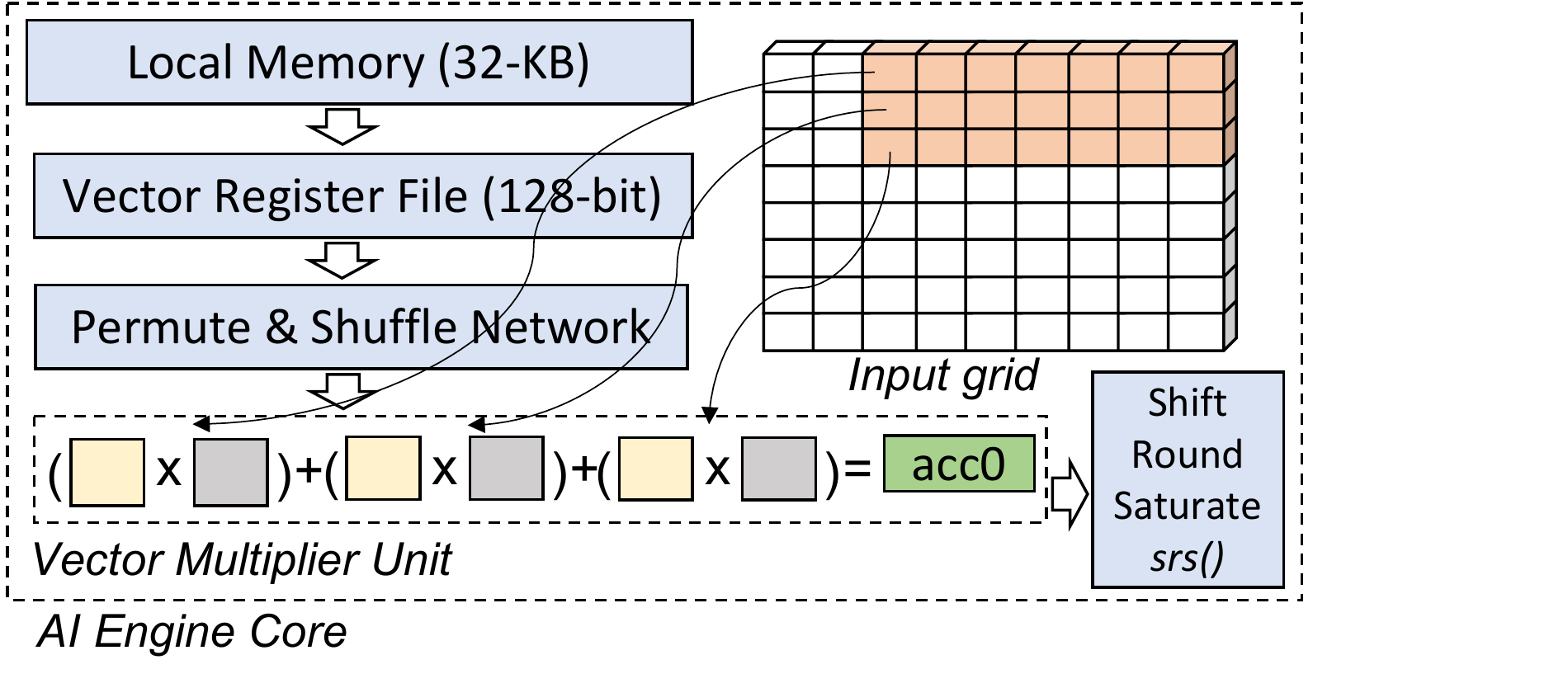}
     \vspace{-0.45cm}
  \caption{Mapping 3-point 2D Jacobi \gcam{(\jacTwoDT)} elementary stencil onto an AIE core. Unlike \hdiff, elementary stencils apply a single stencil pattern throughout the input grid.\label{fig:stencil1d_aie}}
 \end{figure}

Unlike \hdiff, these stencils apply a single stencil pattern throughout the input grid. 
Therefore, such stencils have a higher arithmetic intensity than \hdiff. 
 We illustrate the mapping of elementary stencils on the AIE in 
\fig{\ref{fig:stencil1d_aie}} using an elementary 3-point 2D stencil, i.e., \jacTwoDT. This stencil performs three MAC operations to calculate a single output result. Since the stencil accesses three grid cells in three rows, we tile and load these three rows to an AIE core local memory for computation. Unlike our \hdiff implementation, we do not split the computation of a single elementary stencil into dual-AIE or tri-AIE designs because these elementary stencils perform only a few operations.  

While scaling these elementary stencil designs, a dedicated shimDMA channel is assigned to a specific AIE core; therefore, we enable as many shimDMA channels as the number of AIE cores. This \gommm{dedicated connection} allows us to use the shimDMA bandwidth effectively because each AIE core fetches from an independent shimDMA channel. 

%% file: algorithms/mlir.tex
\begin{lstlisting}[label={lst:hdiffmlir},escapechar=|,caption=MLIR code \gom{example} for \hdiff single-AIE design.]
module @hdiff_single_AIE{
  %t71 = AIE.tile(7, 1) //AIE core  at location 7,1|\label{line:aiecore}|
  %t70 = AIE.tile(7, 0) //shimDMA at location 7,0  |\label{line:shimdma}|
  //setting input and output object FIFOs
  %obj_in = AIE.objectFifo.createObjectFifo(%t70, {%t71}, 5) {sym_name = "obj_in" }: !AIE.objectFifo<memref<i0xi32>> |\label{line:fifoin}|
  %obj_out = AIE.objectFifo.createObjectFifo(%t71, {%t70}, 1){sym_name = "obj_out" } : !AIE.objectFifo<memref<i0xi32>> |\label{line:fifoout}|
   //allocating external DDR buffers
  %ext_buffer_in  = AIE.external_buffer  {sym_name = "ddr_test_buffer_in0"}: memref<in1xi32> |\label{line:extin}|
  %ext_buffer_out = AIE.external_buffer  {sym_name = "ddr_test_buffer_out"}: memref<out1xi32> |\label{line:extout}|
  //point the object FIFOs to the external DDR buffers
  AIE.objectFifo.registerExternalBuffers(%t70, %obj_in : !AIE.objectFifo<memref<i0xi32>>, {%ext_buffer_in}) : (memref<in1xi32>) |\label{line:regin}|
  AIE.objectFifo.registerExternalBuffers(%t70, %obj_out : !AIE.objectFifo<memref<i0xi32>>, {%ext_buffer_out}) : (memref<out1xi32>) |\label{line:regout}|
  //setting AIE core
  %c13 = AIE.core(%t71) { |\label{line:coreinside}|
    scf.for %iv = %lb to %ub step %step {  |\label{line:core_loop}|
      %obj_in_subview = AIE.objectFifo.acquire<Consume>(%obj_in : !AIE.objectFifo<memref<i0xi32>>, 5) : !AIE.objectFifoSubview<memref<i0xi32>>  |\label{line:core_acquire_in_sub}|
      %obj_out_subview = AIE.objectFifo.acquire<Produce>(%obj_out : !AIE.objectFifo<memref<i0xi32>>, 1) : !AIE.objectFifoSubview<memref<i0xi32>> |\label{line:core_acquire_out_sub}|
      func.call @vec_hdiff(%obj_in_subview,%obj_out_subview) : ( memref<i0xi32>,  memref<i0xi32>) -> () |\label{line:core_func_call}|
      AIE.objectFifo.release<Consume>(%obj_in : !AIE.objectFifo<memref<i0xi32>>, 1)  |\label{line:core_release_in_sub}|
      AIE.objectFifo.release<Produce>(%obj_out : !AIE.objectFifo<memref<i0xi32>>, 1) |\label{line:core_release_out_sub}|
    }
    AIE.objectFifo.release<Consume>(%obj_in : !AIE.objectFifo<memref<i0xi32>>, 4) |\label{line:core_release_in_remaing_sub}|
    AIE.end
  } { link_with="hdiff.o" }
}

\end{lstlisting}

%% file: sections/04-evaluation.tex
\section{Evaluation}

\subsection{Experimental \gom{Methodology}}  
We evaluate our accelerator designs for  \hdiff and five elementary stencils   in terms of performance on \gom{an} AMD-Xilinx Versal VCK190~\gom{\cite{vck190}} featuring Versal AI Core XCVC1902-2MSEVSVA2197 ACAP~\cite{aiexcvc1902}. 
\ics{We use hardware locks to measure the end-to-end performance of AIE cores.}
\gomm{Table~\ref{tab:systemparameters} provides our system parameters \gommm{and hardware configuration}.}  
We integrate our designs into the open-source MLIR-AIE framework~\cite{xilinxMLIR}. We use 32-bit precision for all our implementations as used in production by
the Swiss Federal Office of Meteorology and Climatology
(MeteoSwiss)~\cite{osuna2020dawn}. 
\gcam{We generate highly optimized baseline code for CPU and GPU\gommm{-based platforms} using \gommm{the} official MeteoSwiss stencil benchmark suite~\cite{meteobench}.}
We run all our experiments using a $256 \times 256 \times 64$ grid domain similar to the COSMO weather
prediction model~\cite{cosmo_knl}. \gom{We open-source all our code at \MYhref[cyan]{https://github.com/CMU-SAFARI/SPARTA}{https://github.com/CMU-SAFARI/SPARTA}.}

\input{tables/system}

\subsection{Performance Analysis \gom{Results}}
\begin{figure}[h]
\centering
\includegraphics[width=\linewidth,trim={0cm 0cm 0cm 1.7cm},clip]{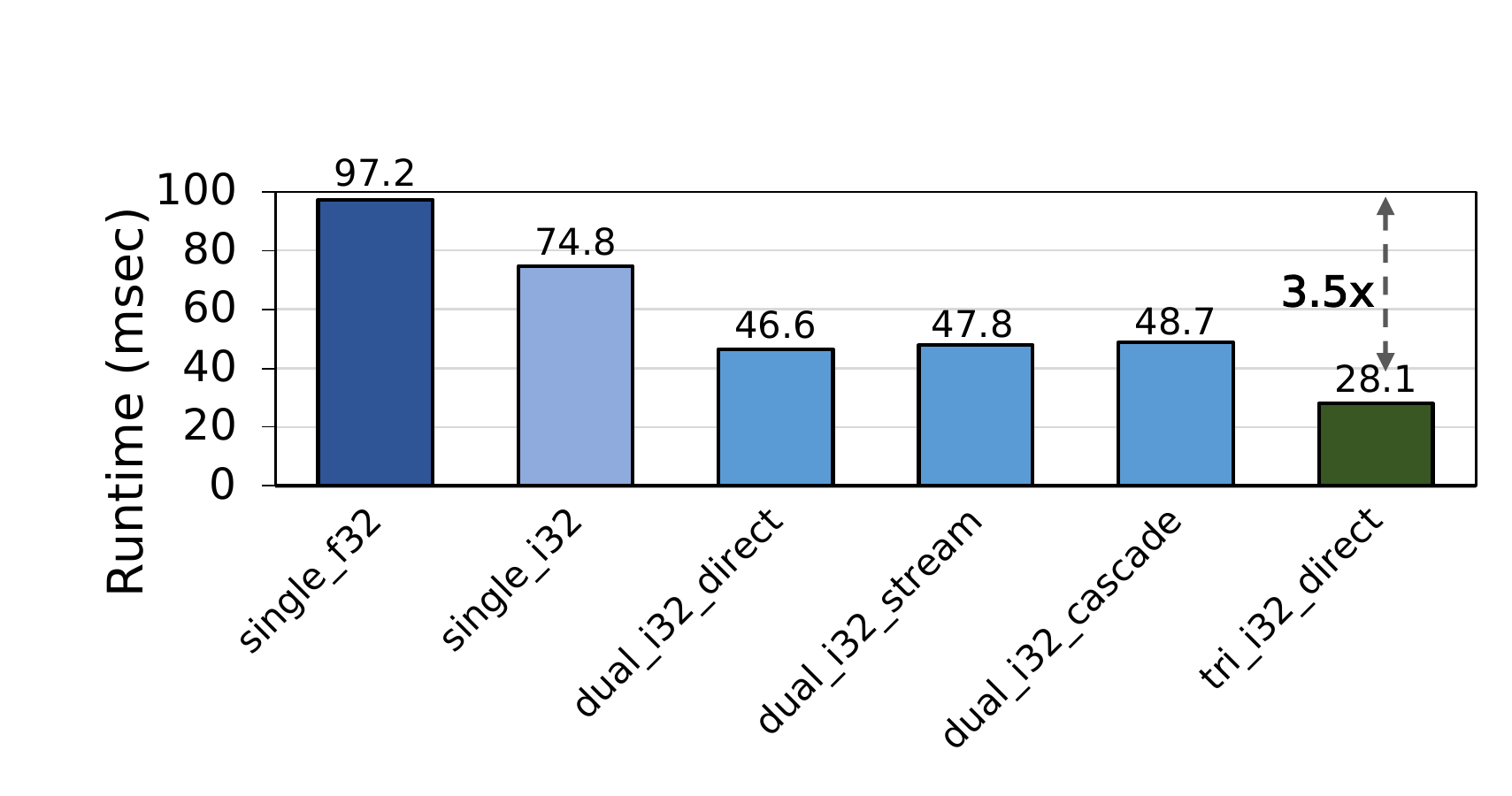} 
\vspace{-0.6cm}
\caption{Runtime (ms) \gom{of} \hdiff using different designs.}
\label{fig:hdiff_design}
\end{figure}

\head{Single-AIE and Multi-AIE Design\gom{s}} 
Figure~\ref{fig:hdiff_design} shows the execution time for  \hdiff~\gom{on the AMD-Xilinx Versal platform} using \saie and \maie designs.

We make the following three observations. First, \gom{the} \taie design (\texttt{tri\_i32\_direct}) improves performance by 3.5$\times$ compared to the \saie design with floating-point implementation (\texttt{single\_f32}). This performance improvement is due to pipelining and balancing of \lap and \flx computation\gom{s} using three separate cores, \gcam{in addition to using the fixed-point datapath}. Second, \gom{the} \maie designs \gom{(i.e., \texttt{\gomm{dual}\_i32\_direct}, \texttt{\gomm{dual}\_i32\_stream, and \texttt{\gomm{dual}\_i32\_cascade}})} improve performance by 1.94$\times$-2.07$\times$ compared to \gom{the} \saie design using fixed-point implementation, depending upon the data forwarding interface. \gcam{The cascading interface provides the lowest improvements as it} should only be used if the \gom{next} \aie~\gom{core} would continue calculating while the data is in the accumulator register. \gom{Otherwise, if the data needs to be returned to the vector register \gomm{from} the accumulator register,} it would entail additional \srs latency of four clock cycles.  Third, using a floating-point datapath (\texttt{single\_f32}) instead of a fixed-point datapath (\texttt{single\_i32}) leads to $1.3\times$ higher execution time because: (1) there are no floating-point accumulator registers due to which we need to use vector registers \gom{for} both  vector operations and data accumulation, which leads to frequent register spilling, and (2) floating-point execution \gom{has longer latencies}. We conclude \gom{that} splitting computation across multiple \gom{AIE} cores 
 leads to a balance of comput\gommm{ation}, memory, and communication requirements, \gommm{enabling} higher performance.
 
\begin{figure}[h]
\centering
\includegraphics[width=\linewidth,trim={0cm 0.6cm 0cm 0cm},clip]{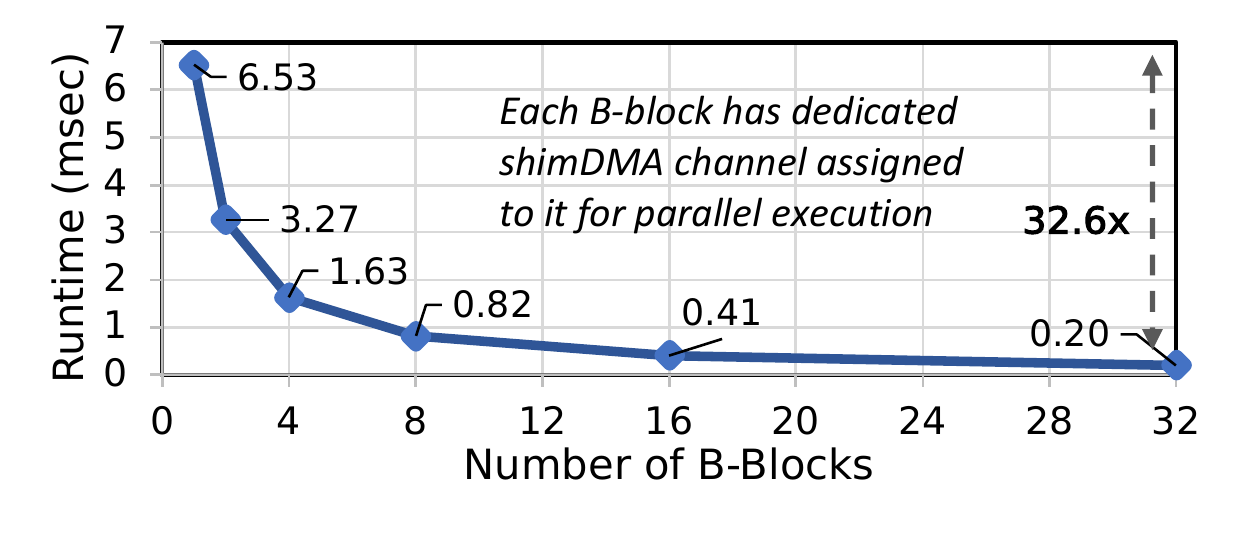} 
\caption{Runtime time (ms) \gom{of} \hdiff with different number of B-blocks.}
\label{fig:hdiff_scale_b_block}
\end{figure}

\head{Scaling Analysis} 
We provide the execution time of \hdiff ~\gom{on the AMD-Xilinx Versal platform using \gommm{the} maximum possible AIE cores} in Figure~\ref{fig:hdiff_scale_b_block}. We scale the number of B-blocks from 1 to the maximum number that we can accommodate \gom{using the available} shimDMA channels. The maximum number of B-blocks is 32 while using 16 shimDMA cores that utilize 384 AIE cores (16 shimDMAs $\times$ 2 B-blocks per shimDMA $\times$ 12 AIE cores per B-block). 

Based on our analysis, we make three key observations. First, the full-blown \mech design (with the maximum number of B-blocks) provides 32.6$\times$ higher performance than a single B-block design. Second, a single B-block is 4.3$\times$ faster than our single tri-AIE design (\texttt{tri\_i32\_direct} in Figure~\ref{fig:hdiff_design}). \gom{A single B-block processes different offsets of output data by broadcasting input data to all the lanes in the B-block} while using the same shimDMA read/write channel. \gom{Broadcasting data to multiple AIE cores at once \gomm{reduces} the communication overhead associated with point-to-point data transfers.}  Third, the performance of \hdiff scales linearly with the number of B-blocks, as all B-blocks have a dedicated shimDMA channel assigned to them. \gom{Having a dedicated shimDMA channel} avoids contention in the memory channels. We conclude \gom{that} our B-block design provides an efficient approach to scale \hdiff implementation across  multiple processing cores while having limited external memory interfaces in a spatial computing system.






\subsection{Performance Comparison \gom{to the State-of-the-Art}}

Table~\ref{tab:compare_works} shows the performance comparison of \mech with state-of-the-art \hdiff implementations. \gom{We mention their reported achieved performance (\texttt{Perf. (GOp/s})) and calculate the percentage of achieved peak roofline performance (\texttt{Arch. Roof. (\%)}) on their respective platforms.} We make use of 384 AIE cores in \mech. We make the following two observations. First, \mech provides the highest performance with only 25.6 GB/s of external memory bandwidth compared to all the other implementations, outperforming  them by 1.2$\times$-31.1$\times$ \gomm{in terms of} GOp/s. Compared to the state-of-the-art CPU, GPU, and FPGA implementations of \hdiff, \mech achieves \hdiffcpu, \hdiffgpu, and \hdifffpga higher performance, respectively. Second, \mech achieves the highest peak roofline performance of 32.2\%, while state-of-the-art implementations can reach only 1.6\%-13.5\% of the peak theoretical performance of a platform. 
This low peak performance is because weather stencils have several elementary stencils cascaded together with data dependencies that lead to complex \gom{and irregular}  memory access patterns. \gcam{Using AIE-Power Estimator~\cite{aiepower}, we estimate the power of \mech as $\sim$23.6 Watts (i.e., energy-efficiency of 42.19 GOps/Watt). Compared to the state-of-the-art FPGA design~\cite{singh2020nero} with 17.34 GOps/Watt, \mech is 2.43$\times$ more energy-efficient.} We conclude \gom{that} \mech provides both higher performance and \gom{higher energy} efficiency compared to all  state-of-the-art implementations \gomm{on CPU, GPU, and FPGA-based computing systems.}
\input{tables/relatedWork.tex}

\subsection{Elementary Stencils Evaluation}

\begin{figure}[h]
\centering
\includegraphics[width=\linewidth,trim={0cm 1.9cm 0cm 0cm},clip]{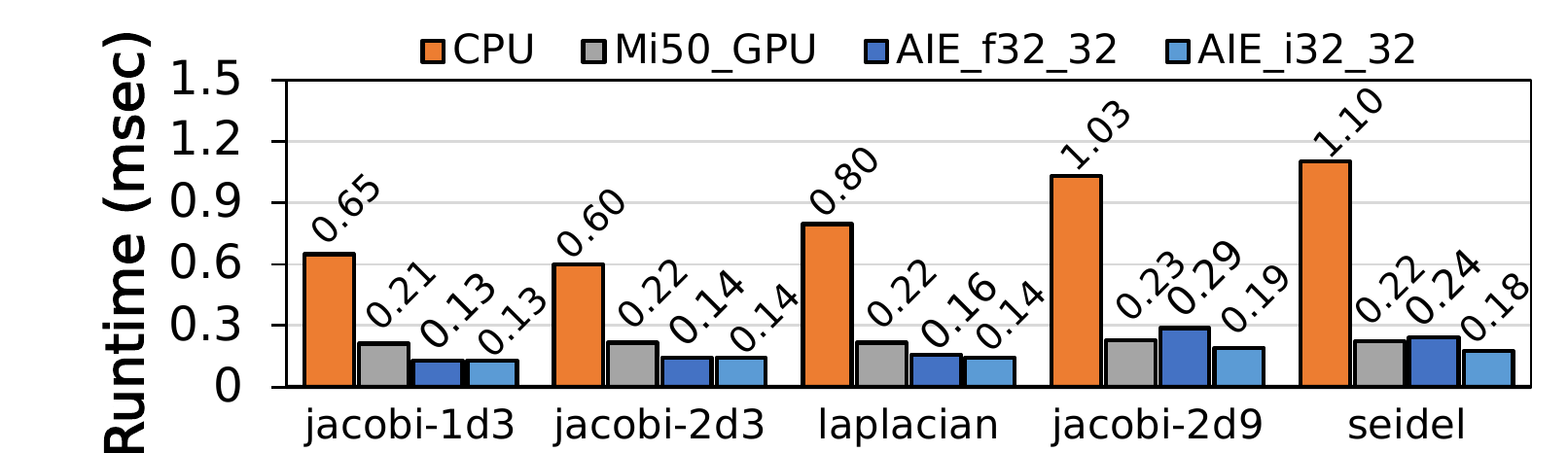} 
\vspace{-0.3cm}
 \caption{Runtime (ms) comparison for elementary stencils on AIE with CPU- and GPU-based platforms.} 
 \label{fig:elementary_design}
 \end{figure}

 Figure~\ref{fig:elementary_design}
shows the execution time comparison for elementary stencils on CPU, GPU, and \gcam{AIE} platform\gom{s}. 

We scale computation to 32 AIE cores for elementary stencils as we develop a single AIE core design that can only access 32 shimDMA channels in parallel. We make the following two key observations. First, \aie-based design outperforms CPU-based implementation (\texttt{CPU}) by 4.5$\times$-6.2$\times$  and 1.5$\times$-4.1$\times$ for fixed-point  (\texttt{AIE\_i32\_32}) and floating-point implementations (\texttt{AIE\_f32\_32}), respectively. Second, AIE, even with only 32 \aies, provides competitive performance compared to our baseline GPU system (\texttt{Mi50\_GPU})~\gom{\cite{mi50}}. We conclude that \gom{the} AIE \gom{platform} provides efficient processing of stencil-based workloads by \gom{effectively} balancing comput\gommm{ation} and data movement.

%% file: tables/system.tex
\vspace{0.25cm}
\begin{table}[h]
\caption{System parameters and hardware configuration for the baseline CPU, GPU, and AMD-Xilinx Versal.}
    \label{tab:systemparameters}
      \begin{center}
    \renewcommand{\arraystretch}{1}
\setlength{\tabcolsep}{2pt}
    \resizebox{1\linewidth}{!}{%
\begin{tabular}{|c|c|}
    \hline
    \textbf{\gommm{CPU}} 
&
\begin{tabular}[c]{@{}l@{}}
         AMD EPYC 7742~\cite{amdEPYC} \\ @2.25GHz, 4-way SMT~\cite{smt}
     \end{tabular}  \\   
     \hline
     \textbf{Cache-Hierarchy}&32$\times$32 KiB L1-I/D, 512 KiB L2, 256 MiB L3 \\
     \hline
     \textbf{System Memory}&4$\times$32GiB RDIMM DDR4 2666 MHz~\cite{rdimm} PCIe 4.0 $\times$128 \\
    \hline
     \textbf{OS details}& 
      \begin{tabular}[c]{@{}l@{}}
         Ubuntu 21.04 Hirsute Hippo~\cite{ubuntu},\\ GNU Compiler Collection (GCC) version 10.3.0~\cite{gnu}
     \end{tabular}  \\
      \hline
      \hline
      \textbf{GPU} &    \begin{tabular}[c]{@{}l@{}}
     AMD Radeon Instinct™ MI50~\cite{mi50} 3840 Stream Processors\\ 32GB HBM2 PCIe 4.0 $\times$16, ROCm version 5.1.1~\cite{rocm511}  
         
     \end{tabular}  \\ \hline

     \hline
      \hline
     \textbf{Versal} &    \begin{tabular}[c]{@{}l@{}}
          XCVC1902-2MSEVSVA2197 AI Core~\cite{aiexcvc1902}\\ 400$\times$\aies @1GHz,
         8GB DDR4 DIMM~\cite{xilinxdimm},\\  Dual-Core Arm Cortex-A72~\cite{armOnAIE}
     \end{tabular}  \\ \hline
\end{tabular}
}
  \end{center}
 \end{table}

%% file: tables/relatedWork.tex
\begin{table*}[h]
  \caption{Overview \gom{and performance} of the state-of-art \gcam{\hdiff} implementations. For each work, we mention the  DRAM memory technology (Mem. Tech.), theoretical peak floating-point  performance (Peak Perf. (TFLOPS)), available peak memory bandwidth (Peak B/W (GB/s)), 
 achieved performance (Perf. (GOp/s)), and the percentage of achieved peak roofline performance (A\gom{r}ch. Roof. (\%)). }  
    \label{tab:compare_works}
      \begin{center}
      \small
\resizebox{1\linewidth}{!}{%
\begin{tabular}{c|c|c|l|lH|l|l|lH|HHl|l}
\textbf{Stencil} & Work & Year & Platform&Device & Tech. node & Mem. Tech. & Peak Perf. (TFLOPS) & Peak B/W (GB/s)   & Freq. (MHz)& Logic Util. & Mem. Util. & Perf. (GOp/s) & A\gom{r}ch. Roof. (\%)   \\
\hline
\hdiff & \cite{narmada} & 2019&FPGA & XCVU3P~\gom{\cite{vu37p}} & TSMC 16FF+ & DDR4&0.97 &25.6 &200 & 64.5\% & 64.1\% & 129.9& 13.4\% \\%
\hdiff & \cite{de2021stencilflow} & 2021&CPU &   Xeon E5-2690V3~\gom{\cite{xeon_e5_2690}} & Intel 14nm FinFet & DDR4& 0.24 &68.0&292-317 & 26.0\% & 20\% & 32.0& 13.0\% \\ %
\hdiff & \cite{singh2020nero} & 2021&CPU & POWER9~\gom{\cite{POWER9}} & TSMC 16FF+ & DDR4 & 0.49 & 110.0 & 58.5& 12.5\% & 52\% & 58.5& 11.8\%  \\%
\hdiff & \cite{de2021stencilflow} & 2021&GPU &   V100~\gom{\cite{v100}} & Intel 14nm FinFet & HBM2& 14.1 &900.0&292-317 & 26.0\% & 20\% & 849.0& 6.1\% \\ %
\hdiff & \cite{de2021stencilflow} & 2021&FPGA &  Stratix 10~\gom{\cite{intel_stratix10}} & Intel 14nm FinFet & DDR4& 9.2 &76.8&292-317 & 26.0\% & 20\% & 145.0& 1.6\% \\ %

\hdiff & \cite{singh2020nero} & 2021&FPGA & XCVU37P~\gom{\cite{vu37p}} & TSMC 16FF+ & HBM & 3.6 & 410.0 & 250& 12.5\% & 52\% & 485.4& 13.5\%  \\%
\textbf{\hdiff} & \textbf{\mech} & \textbf{2023}&\textbf{AIE} & \textbf{XCVC1902}~\gom{\cite{vck190}}  & TSMC 16FF+ & \textbf{DDR4} & \textbf{3.1} & \textbf{25.6 }& 1000& 96\%$^\dagger$ & 52\% & \textbf{995.7} & \textbf{32.2\%}  \\%

\hline
\end{tabular}
}
\end{center}
\end{table*}



%% file: sections/05-discussion.tex
\section{Discussion}


Spatial computing systems, with their high comput\gom{ation} density through a large number of processing cores and customizab\gom{ility using a configurable} interconnect\gomm{ion} network, offer significant \gom{promise to improve performance over modern multicore and heterogeneous systems}. We present six key \gomm{takeaways} that we distill from our  experimental \gomm{design, analysis, and} characterization \gomm{on} \gom{a} \gcam{state-of-the-art spatial computing system, i.e.,} \gom{the} AMD-Xilinx Versal AI Engine architecture. 
\begin{itemize}[leftmargin=*, noitemsep, topsep=0pt]
\item \taway{1}{\gcam{Workload balancing \gomm{and parallelization} across} comput\gomm{ation} resources \gomm{is important}} Spatial architectures provide a high degree of parallelism. To fully leverage this parallelism, it is essential to design the application such that computation is partitioned into parallelizable tasks. Our experimental results utilizing a single AIE core-based design demonstrate that a lack of optimization in compute utilization can result in suboptimal resource utilization, leading to up to 2.77$\times$ lower performance compared to a \maie design using the same datapath.

\item\taway{2}{Use broadcasting} \gom{We} exploit the broadcast feature of AIE that allows transferring data to multiple cores at the same time, \gom{which} reduces data transfer overhead\gom{s}. \mech provides 4.3$\times$ higher throughput per shimDMA channel by broadcasting data using our B-block-based approach.

\gcam{
\item \taway{3}{\gomm{Ease of} programm\gomm{ing is beneficial}} AIE-cores are software-programmable \gom{and} can adapt to different applications without physical reconfiguration. Single-core AIE programming, similar to GPU programming, uses C/C++ constructs with intrinsics. However, the use of the MLIR framework for AIE programming allows developers to focus on high-level logic while handling lower-level details of code generation and optimization for different hardware targets. \gomm{Such programming ease was helpful in our development and design space exploration.} 
}

\item\taway{4}{\gcam{Use \gomm{the proper} datatype with the highest comput\gommm{ation} density}} Using a fixed-point datapath can be more flexible and efficient for some applications. For example, \mech using \gom{the} fixed-point datapath achieves 1.3$\times$ higher performance compared to the floating-point datapath. 
The fixed-point datapath offers advantages such as: (1) support for dedicated accumulator registers\gom{, which allows} for more efficient operations, (2) faster processing time, and (3) support for a wider range of precision options.

\item\taway{5}{Use multiple accumulator/vector registers for parallel execution} \aie provides 8$\times$256-bit vector and 4$\times$384-bit accumulator registers. \gom{We use} multiple registers \gom{to}: (1) allow \gom{dependence}-free execution and (2) hide the latency of computation (e.g., \gomm{since} floating-point MAC has a latency of 2 cycles, we can schedule two MAC operations in a pipeline\gom{d} manner). 

\gcam{
\item\taway{6}{Assist compiler in better scheduling}  \gomm{In order to} assist the compiler in \gomm{performing} better scheduling, \gom{we} restructure the code to reduce dependencies and improve parallelism. \gom{We achieve this} by: \textbf{(a) interleaving memory and arithmetic operations:} \gom{by} rearranging loads, MAC/MUL, and store operations\gommm{,} \gom{we} maximize the use of the VLIW pipeline and avoid NOP instructions; \textbf{(b) avoiding register spilling}: \gom{we} avoid having too many \gom{active} registers (both vector and accumulators) since hardware can only support a limited number of registers (8$\times$256-bit vector and 4$\times$384-bit accumulator registers). Instantiating more variables than available registers leads to register spilling \gomm{(i.e.,} causes data to spill onto stack memory\gomm{, which reduces performance)}; \textbf{(c)  shortening dependency length of a loop:} \gom{we} load data for the first iteration outside of the loop body and for the next iteration at the end of a loop to shorten the dependency length of a loop.}

\end{itemize}

%% file: sections/06-related.tex
\section{Related Work}
To our knowledge, this is the first work to evaluate the benefits of \gom{using} a cutting-edge spatial computing system for \gom{the horizontal diffusion stencils commonly used in} real-world weather  \gomm{and climate} \gom{modeling workloads}~\gomm{\cite{gysi2015modesto,cosmo_knl,de2021stencilflow,palmer1990european,mcclung2016global,hurrell2013community,watanabe2011miroc}}. In this section, we describe \gomm{various} other related works on accelerating stencil processing.

Stencil computation represents an important class of scientific workloads \gom{used in} many high-performance computing applications,  including computational fluid dynamics~\cite{huynh2014high}, image processing~\cite{hermosilla2008non}, weather prediction modeling~\cite{doms1999nonhydrostatic}, \gom{seismic imaging \cite{mcmechan1983migration}, electromagnetic simulations~\cite{taflove1988review},} \gomm{heat diffusion~\cite{frigo2007memory}, astrophysics~\cite{balsara2017higher}, quantum computing~\cite{kormann2010error}, cellular automata~\cite{chi2018soda}}, etc. Many past works focus on optimizing only elementary stencil computations~\cite{augustin2009optimized,datta2009optimization,datta2008stencil,de2009introducing,datta2009auto,dursun2009core,dursun2009multilevel,kamil2005impact,krishnamoorthy2007effective,li2004automatic,meng2009performance,micikevicius20093d,waidyasooriya2019multi,sano2014multi,gan2017solving,van2019coherently,denzler2023casper,
li2019pims,nguyen20103,stengel2015quantifying,fuhrer2018near,armejach2018stencil,yantir2020efficient,wester2014deriving,christen2011patus,olschanowsky2014study,brandvik2010sblock,phillips2010implementing,szustak2013using,wang2017comprehensive,sohrabizadeh2022autodse,reggiani2021enhancing,koraei2019dcmi,tian2022sasa,chi2018soda}. However, \emph{compound stencils}, which are made up of a series of elementary stencils that perform element-wise computations on a full 3D grid, are of great significance in real-world applications. \gom{Some past} works \gom{use} commodity devices that have a large market volume \gomm{to accelerate compound stencil computations}. \gom{For example,} Bianco et al.~\cite{bianco2013gpu} optimize the COSMO Model for GPU processing\gommm{,}  while Thaler et al.~\cite{cosmo_knl} port COSMO to a many-core system. 
More recent \gom{works} use FPGAs to accelerate compound stencils~\cite{singh2020nero,narmada,singh2019low,de2021stencilflow}. 
However, taking full advantage of 
FPGAs to accelerate a workload is not trivial~\cite{singh2022leaper,singh2021modeling}. Compared to CPUs or GPUs, {an} FPGA must exploit an order of magnitude more parallelism in a target workload to compensate for the lower clock frequency~\cite{diamantopoulos2020agile,jun2015bluedbm, cali2022segram,alser2022molecules, narmada,10.14778/3137765.3137776,jiang2020}. We observe that real-world weather stencils \gom{have high hardware resource utilization and low peak performance because these stencils} cannot \gom{easily} leverage the bit-level \gcam{flexibility} available on FPGAs. \gom{O}ur work provides the first comprehensive evaluation of a state-of-the-art spatial architecture \gom{(that consists of an array of cores with a customizable interconnect)} for real-world stencil computation.
 

%% file: sections/07-conclusion.tex
\section{Conclusion}

Real-world climate and weather \gom{modeling applications} require the use of complex compound stencil kernels that consist of a combination of different stencil \gom{computation} \gcam{patterns}. Horizontal diffusion is a fundamental compound kernel, which is an important part of many  climate and weather prediction models. \gcam{State-of-the-art implementations of horizontal diffusion are memory bound with limited performance on current CPU, GPU, and FPGA-based computing systems.} 
We propose \mech,  a novel spatial accelerator for horizontal diffusion weather stencil
computation. 
This work
is the first to propose and evaluate a real spatial architecture \gom{(i.e., AMD-Xilinx Versal AI Engine)} \gom{specifically}
for weather stencil computation, as opposed to commonly used
machine learning-based workloads. 
Our results on a real spatial computing system demonstrate that \mech outperforms all state-of-the-art CPU, GPU, and FPGA implementations. We conclude that \gom{modern} spatial architectures have the potential to provide \gom{both} specializ\gom{ation and ease of programm\gomm{ing}}, surpassing the limitations of traditional platforms for real-world weather \gom{and climate modeling}. We open-source all our code at \MYhref[cyan]{https://github.com/CMU-SAFARI/SPARTA}{https://github.com/CMU-SAFARI/SPARTA} to facilitate reproducibility and ease of use \gom{as well as enable future research}. We hope this work \gom{provides} an important step \gom{towards} using spatial computing systems for complex weather \gom{and climate} modeling \gom{workloads}.
